\documentclass[lettersize,journal]{IEEEtran}
\usepackage{amsmath,amsfonts,amssymb,amsthm}
\usepackage{algorithmic}
\usepackage{algorithm}
\usepackage{array}
\usepackage[caption=false,font=footnotesize,labelfont=rm,textfont=rm]{subfig}
\usepackage{textcomp}
\usepackage{stfloats}
\usepackage{url}
\usepackage{verbatim}
\usepackage{graphicx}
\usepackage{cite}
\usepackage{xcolor}
\usepackage{booktabs}
\usepackage{makecell} 
\begin{document}

\title{Two-Timescale Antenna Positioning for Pinching-Antenna System Based on Statistical Channel Information}

\author{Jingze Ding, \IEEEmembership{Graduate Student Member, IEEE},
		Zijian Zhou, \IEEEmembership{Member, IEEE}, \\
		Bingli Jiao, \IEEEmembership{Senior Member, IEEE},
		and Rui Zhang, \IEEEmembership{Fellow, IEEE}
\thanks{This work is supported in part by National University of Singapore under Research Grants A-8003646-00-00 and A-8003676-00-00, and in part by China Scholarship Council. An earlier version of this paper was accepted in part at the IEEE GLOBECOM 2026 \cite{conf}. \textit{(Corresponding author: Rui Zhang.)}}
\thanks{Jingze Ding is with the School of Electronics, Peking University, Beijing 100871, China, and also with the Department of Electrical and Computer Engineering, National University of Singapore, Singapore 117583 (e-mail: djz@stu.pku.edu.cn).}
\thanks{Zijian Zhou is with the School of Science and Engineering, The Chinese University of Hong Kong, Shenzhen 518172, China (e-mail: zijianzhou@link.cuhk.edu.cn).}
\thanks{Bingli Jiao is with the School of Computing and Artificial Intelligence, Fuyao University of Science and Technology, Fuzhou 350109, China, and also with the School of Electronics, Peking University, Beijing 100871, China (e-mail: jiaobl@pku.edu.cn).}
\thanks{Rui Zhang is with the Department of Electrical and Computer Engineering, National University of Singapore, Singapore 117583 (e-mail: elezhang@nus.edu.sg).}
}
\maketitle

\begin{abstract}
	This paper introduces a two-timescale antenna position optimization framework for the pinching-antenna system (PASS). The proposed framework exploits statistical channel state information (CSI) to determine large-timescale pinching-antenna (PA) positions over the entire waveguide, and utilizes instantaneous CSI to refine PA positions at the small timescale within a local region. This approach effectively captures the benefits of large-timescale PA positioning for long-term system performance and small-timescale PA movement for real-time adaptation to fast channel fluctuations. For the single-user case, we derive closed-form solutions for the optimal large-timescale PA positions, and further analyze a performance upper bound together with the conditions for achieving this bound, which guide the small-timescale position refinement. For the multi-user case, an upper bound on the minimum rate among all users is derived. To maximize their minimum rate, we develop an element-wise search algorithm for large-timescale PA positioning and a projected gradient ascent method for small-timescale position refinement. Simulation results show that the proposed scheme closely approaches the derived bounds and consistently outperforms conventional schemes by reducing path loss via large-timescale PA positioning and suppressing inter-user interference through small-timescale local movement.
\end{abstract}
\section{Introduction}
As a cornerstone of modern wireless communications, multiple-input multiple-output (MIMO) technology has played a pivotal role in improving the spectral efficiency and link reliability of wireless communication systems \cite{MIMO1}. Building on conventional MIMO, a series of flexible MIMO technologies have recently emerged, such as movable/fluid antennas \cite{MA1,MA2,MA3,6DMA1} and polarization-reconfigurable antennas \cite{PRA1,PRA2,PRA3,PRA4}, which introduce additional spatial and electromagnetic degrees of freedom (DoFs) to more efficiently allocate communication resources based on diverse application requirements. However, despite their demonstrated ability to mitigate small-scale channel fading and adapt to diverse statistical channel conditions \cite{MA_small1,MA_small2}, these flexible MIMO techniques are still fundamentally constrained by centralized antenna deployments, where antennas are centrally mounted at the base station (BS) or access point (AP). As a result, they cannot directly address large-scale path loss that is dominated by the signal propagation distance. Moreover, these flexible MIMO systems generally lack hardware flexibility, since once deployed, adding or removing antennas is practically difficult and inconvenient.
\subsection{Pinching-Antenna System}\label{pass}
To overcome the above limitations, NTT DOCOMO recently introduced the pinching-antenna system (PASS) as a new flexible antenna architecture \cite{PASS1}. In PASS, dielectric waveguides are used as transmission media, where small dielectric particles, referred to as pinching antennas (PAs), are attached at designated positions to radiate or receive electromagnetic waves \cite{PASS2}. Unlike other flexible MIMO systems, PASS supports a decentralized antenna deployment, since the waveguides can be extended over a large area \cite{PASS3}. By placing PAs in close proximity to users, PASS can establish stable line-of-sight (LoS) channels and effectively reduce the signal propagation distance, thereby improving robustness against blockage and alleviating large-scale path loss \cite{PASS4}. Moreover, PASS enables low-complexity and low-cost implementation by simply adding or removing dielectric particles \cite{PASS5}.

Due to the above advantages, the applications of PASS in wireless communication and sensing systems have recently attracted increasing research attention \cite{PASS_comm1,PASS_comm2,PASS_comm3,PASS_comm4,PASS_ISAC1,PASS_ISAC2,PASS_ISAC3}. In existing studies, PA deployment strategies can be categorized into discrete \cite{PASS_disc1,PASS_disc2}, continuous \cite{PASS_cont1,PASS_cont2}, and semi-continuous \cite{PASS3,PASS_semi1} positioning schemes. 
\subsubsection{Discrete PA Positioning}
For discrete PA positioning, a set of uniformly spaced PAs is pre-installed along the waveguide, and only a small subset is activated when needed \cite{PASS_disc1}. With programmable electrical control, discrete activation can be achieved by regulating the PA-waveguide separation. By varying this separation, the coupling strength can be tuned, thereby enabling PAs to be selectively activated or deactivated \cite{PASS_power}. Such discrete activation enables fast reconfiguration and low hardware complexity without relying on mechanical modules, which is scalable for large-scale PASS design \cite{PASS_disc2}. However, the performance may be limited by the quantization of feasible PA positions, especially when fine-grained spatial adaptation is required.
\subsubsection{Continuous PA Positioning}
For continuous PA positioning, each PA is mounted on a slide track along its associated waveguide and is driven by a dedicated linear actuator \cite{PASS_cont1}, with a position sensor and a real-time controller to support closed-loop position tracking and adjustment. The key advantage of continuous PA positioning is that it provides the highest spatial resolution, which allows more effective exploitation of channel variations. Nevertheless, this flexibility comes at the cost of increased hardware complexity and control overhead \cite{PASS_cont2}. Moreover, frequent mechanically actuated positioning may incur non-negligible actuation latency and energy expenditure \cite{energy}. Hence, continuous PA positioning is best suited for slowly varying environments.
\subsubsection{Semi-continuous PA Positioning}
For semi-continuous PA positioning, the waveguide is divided into multiple segments. Each PA is pre-installed on one segment, and it supports local micro-adjustments within its assigned segment after activation \cite{PASS_semi1}. In practical implementation, each PA can be mounted on a slide track along its assigned segment and driven by a compact stepper motor. By leveraging local refinements within the segment, the semi-continuous PA positioning can retain PA positioning precision for high performance, while reducing both refinement latency and hardware costs compared to the fully continuous PA positioning scheme, thus offering a favorable balance between the high resolution of continuous positioning and the low complexity of discrete activation \cite{PASS3}.
\subsection{Motivations and Contributions}
Although extensive efforts have been devoted to PASS under discrete, continuous, and semi-continuous PA positioning strategies, the majority of existing works adopt a single-timescale formulation and optimize PA positions assuming perfectly known instantaneous channel state information (CSI). Consequently, the following two challenges remain unaddressed. First, the single-timescale formulations optimize PA positions over the entire waveguide based on instantaneous CSI, which leads to a large optimization space. The objective value of the resulting problem exhibits numerous local optima with a large gap to the global optimum \cite{PASS3}. Hence, it is necessary to develop reliable algorithms for PA position optimization to avoid convergence to undesirable local solutions. Second, it is impractical to estimate the instantaneous CSI for arbitrary PA positions due to the prohibitively high complexity, especially in multi-path scenarios with both LoS and non-line-of-sight (NLoS) components \cite{PASS13,PASS_add_nlos}. This is because the near-field effects induced by large-scale PA positioning cause the angles of departure or arrival (AoDs/AoAs) and the complex responses of the channel paths to vary across different PA positions. Therefore, the commonly adopted assumption that perfect instantaneous CSI is available at any PA position mainly serves to characterize an upper bound on system performance, but offers limited insights for practical deployments. These challenges motivate the adoption of a two-timescale optimization framework for PASS design. In particular, the authors of \cite{PASS14} proposed a two-timescale scheme, where the PA positions are treated as long-term variables optimized over the channel coherence time, while the beamforming is optimized in each time slot based on the instantaneous CSI under fixed PA positions. However, PA position optimization across different timescales has not been investigated. As previously discussed, the semi-continuous PA positioning architecture provides a suitable basis for two-timescale PA positioning, as it naturally decouples PA deployment into a two-stage operation that supports both wide-range positioning and high-precision local refinement. This architecture motivates us to investigate a two-timescale PA positioning scheme for PASS to overcome the limitations of conventional single-timescale schemes. At the large timescale, the PA positions are optimized over the entire waveguide based on statistical CSI. At the small timescale, they are locally refined within a small region using only the instantaneous CSI associated with that region. Consequently, instantaneous CSI at arbitrary PA positions is no longer required. The main contributions of this paper are summarized as follows:
\begin{itemize}
	\item First, a statistical channel model is proposed for PASS to capture both the deterministic LoS component and the random variations in the NLoS paths. Based on this model, a two-timescale PA positioning scheme is developed. For the considered PASS-enabled multi-user downlink communication system, the users' maximized minimum (max-min) rate is adopted as the performance metric, and an optimization problem is formulated to jointly optimize the PA positions across the two timescales.
	\item Next, for the special case of a single-user system, the optimal large-timescale PA positions are derived, based on which a performance upper bound and the conditions for achieving this bound are analyzed to provide insights for small-timescale PA position refinement. Subsequently, an efficient parallel search algorithm is developed to obtain the refined PA positions.
	\item Then, for the general multi-user case, an upper bound on the max-min rate is derived under the ideal condition of zero inter-user interference. To approach this bound, an element-wise search algorithm is proposed for large-timescale PA position optimization, while a projected gradient ascent algorithm is developed for small-timescale PA position refinement.
	\item Finally, extensive simulations are conducted to demonstrate the advantages of the proposed two-timescale PA positioning scheme. The results show that large-timescale PA positioning improves user rates by reducing large-scale path loss, while small-timescale PA movement exploits small-scale channel fading to reduce the correlation among users' channel vectors. By jointly exploiting gains at both timescales, the proposed two-timescale PA positioning scheme achieves robust performance under diverse user distributions.
\end{itemize}
\subsection{Organization and Notations}
The rest of this paper is organized as follows. Section \ref{2} introduces the system and channel models, based on which the two-timescale optimization problem is formulated. In Section \ref{3}, a simple single-user scenario is first studied, which is then extended to the general multi-user case in Section \ref{4}. Next, simulation results and discussions are provided in Section \ref{5}. Finally, this paper is concluded in Section \ref{6}.

\textit{Notation:}  $a/A$, $\mathbf{a}$, $\mathbf{A}$, and $\mathcal{A}$ denote a scalar, a vector, a matrix, and a set, respectively. ${\left(  \cdot  \right)^*}$, ${\left(  \cdot  \right)^{T}}$, ${\left(  \cdot  \right)^{H}}$, and $\left|  \cdot  \right|$ denote the conjugate, transpose, conjugate transpose, and absolute value, respectively. $\mathrm{tr}\left(\cdot \right) $ denotes the trace of a matrix. $\angle \left(\cdot \right) $ denotes the phase of a complex number. $\left\|\cdot \right\|_1$ and $\left\|\cdot \right\|_2$ denote the 1-norm and 2-norm of a vector, respectively. $\left\|\cdot \right\|_F$ represents the Frobenius norm of a matrix. $\mathbb{C}$ and $\mathbb{Z}$ denote the sets of complex numbers and integers, respectively. $\mathbb{R}^{M \times N}$ and $\mathbb{C}^{M \times N}$ represent the sets of real and complex matrices/vectors of dimension $M \times N$, respectively. $\mathbb{E}\left(\cdot \right) $ is the expectation of a random variable. $\mathbf{I}_N$ denotes an identity matrix of size $N \times N$. $\mathbf{0}_N$ denotes an $N$-dimensional vector with all elements equal to 0. $\mathcal{CN}\left(0,\sigma^2 \right) $ represents the circularly symmetric complex Gaussian (CSCG) distribution with mean zero and variance $\sigma^2$. $\mathcal{U}\left[a,b \right] $ denotes the uniform distribution within real-number interval $\left[a,b \right]$. $\nabla_{\mathbf{x}}f\left(\mathbf{x} \right) $ denotes the gradient of function $f\left(\mathbf{x} \right)$ with respect to $\mathbf{x}$. $\left[ \mathbf{a}\right]_{n} $ denotes the $n$-th entry of vector $\mathbf{a}$. $\left[ \mathbf{A}\right]_{i,j} $ denotes the entry in row $i$ and column $j$ of matrix $\mathbf{A}$. $\in$, $\sim$, and $\triangleq$ represent ``belongs to'', ``distributed as'', and ``defined as'', respectively.
\section{System Model and Problem Formulation}\label{2}
\begin{figure}[!t]
	\centering
	\includegraphics[width=1\linewidth]{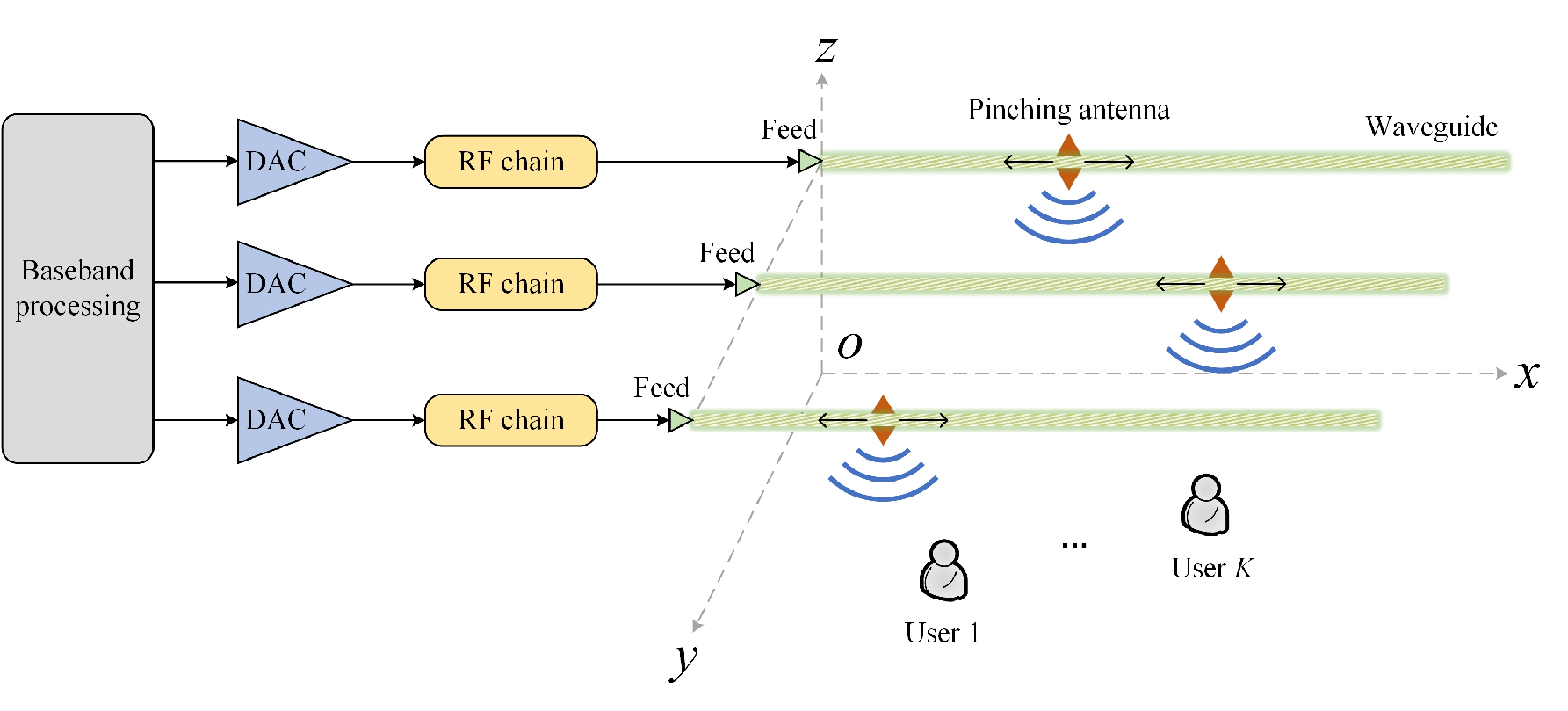}
	\caption{Illustration of the PASS-enabled multi-user downlink communication.}
	\label{system_model}
\end{figure}
As shown in Fig. \ref{system_model}, we consider a multi-user downlink communication system enabled by a PASS. A global Cartesian coordinate system is established to specify the positions of PAs and users. The PASS is equipped with $N$ waveguides, each connected to a dedicated RF chain. A single PA is flexibly pinched along each waveguide\footnote{This paper adopts one PA per waveguide to focus on characterizing and exploiting the performance gains brought by the proposed two-timescale antenna positioning in PASS. The system model and the proposed algorithms can be extended to multiple PAs per waveguide by including all PA positions.}, and the $N$ PAs jointly serve $K$ single-antenna users. Both waveguides and PAs are installed at a fixed height $h$. The users are distributed within a predefined two-dimensional (2D) rectangular area of size $S_\mathrm{x} \times S_\mathrm{y}$. The position of user $k$ ($1 \le k \le K$) is given by $\mathbf{u}_k = \left[ X_k, Y_k, Z_k \right]^T \in \mathbb{R}^{3 \times 1}$. The position of PA $n$ ($1 \le n \le N$) is denoted by ${\mathbf{p}_{n}} = {\left[ {{x_{n}},y_n,{h}} \right]^T} \in \mathbb{R}^{3 \times 1}$, where $x_{n}$ ($0 \le x_{n} \le S_\mathrm{x}$) denotes the adjustable location of this PA along the $x$-axis and $y_n$ ($0 \le y_n \le S_\mathrm{y}$) denotes the $y$-axis coordinate of the feed point for waveguide $n$. To clearly describe the two-timescale position adjustments of the PAs, $x_n^\circ$ is defined as the large-timescale PA position, based on which the small-timescale PA position $x_n$ can vary freely over the small-scale movement region $\mathcal{X}_{n}\triangleq\left[x_n^\circ-{D _{n}}/{2},x_n^\circ+{D _{n}}/{2}\right]  $, i.e., $x_n \in \mathcal{X}_{n}$, $\forall n$, where $D _{n}$ denotes the length of this region. To guarantee that the PA does not exceed the waveguide length during small-timescale position refinement, $x_n^\circ$ is constrained to the large-scale positioning region $\mathcal{X}_{n}^\circ\triangleq\left[{D _{n}}/{2},S_\mathrm{x}-{D _{n}}/{2}\right]  $, i.e., $x_n^\circ \in \mathcal{X}_{n}^\circ$, $\forall n$.
\subsection{Statistical Channel Model}
The PASS channel consists of in-waveguide propagation and free-space propagation. Define $n_\mathrm{e}$ as the effective refractive index of the dielectric waveguide. For in-waveguide propagation, the path response from the feed point of waveguide $n$ to its associated PA is given by
\begin{equation}
	{g_n}\left( {{x_n}} \right) = {\rho_{n}e^{ \jmath\frac{{2\pi }}{{{\lambda _\mathrm{W}}}}{x_{n}}}},
\end{equation}
where $\lambda _\mathrm{W}={\lambda}/{n_\mathrm{e}}$, with $\lambda$ denoting the free-space carrier wavelength. $\rho_{n}$ denotes a power-scaling factor of the PA. For free-space propagation, the Rician channel model is adopted. Assume that there are $L_{k,n}$ propagation paths from the PA $n$'s large-timescale position $x^\circ_n$ to user $k$, whose angles of departure (AoDs) and path responses are denoted by $\theta _{k,n}^l \in \left[0,\pi \right] $ and $b_{k,n}^l \in \mathbb{C}$ ($1\le l \le L_{k,n}$), respectively. When PA $n$ moves within the small-scale region $\mathcal{X}_{n}$, the AoDs $\{\theta _{k,n}^l\}_{l=1}^{L_{k,n}}$ and the amplitude of the complex coefficient for each channel path between PA $n$ and each user do not change, while only the phase of the channel path varies with the PA position. This is because the size of the small-scale movement region is much smaller than the signal propagation distance, and thus the far-field condition is satisfied. Therefore, the free-space channel from PA $n$ to user $k$ is given by
\begin{equation}\label{hkn}
	h_{k,n}\left(x_n \right)=\mathbf{b}_{k,n}^H\mathbf{f}_{k,n}\left( x_n\right), 
\end{equation}
where $\mathbf{b}_{k,n} \in \mathbb{C}^{L_{k,n} \times 1}$ and $\mathbf{f}_{k,n}\left( x_n\right)\in \mathbb{C}^{L_{k,n} \times 1}$ denote the path response vector and the phase differences of the $L_{k,n}$ paths, which are respectively given by
\begin{align}
	\mathbf{b}_{k,n}&=\left[b_{k,n}^1,\ldots,b_{k,n}^{L_{k,n}} \right]^T ,\\
	\mathbf{f}_{k,n}\left( x_n\right)&=\left[e^{{\jmath}\frac{2\pi}{\lambda}\left( x_n-x_n^\circ\right)\cos{\theta _{k,n}^1} }, \ldots, e^{{\jmath}\frac{2\pi}{\lambda}\left( x_n-x_n^\circ\right)\cos{\theta _{k,n}^{L_{k,n}}} } \right]^T.
\end{align}
Let $\beta_0$ and $\alpha$ denote the reference channel gain at 1 meter (m) and the path loss exponent, respectively. Define $d_{k,n}\triangleq\left\|\mathbf{p}_{n}^\circ-\mathbf{u}_k\right\|_2$ and $\kappa_{k,n}$ as the distance and the Rician factor from the large-timescale PA position $x_n^\circ$ to user $k$, respectively, where $\mathbf{p}_{n}^\circ = \left[ x_n^\circ,y_n,h\right]^T \in \mathbb{R}^{3 \times 1} $. Under a fixed $x_n^\circ$, the AoD and path response for the LoS path are deterministic and respectively given by 
\begin{align}
	\theta _{k,n}^1 &= \arccos{\frac{X_k-x_n^\circ}{d_{k,n}}},\forall k,n, \\
	b_{k,n}^1&=\sqrt{\frac{\beta_0d_{k,n}^{-\alpha}\kappa_{k,n}}{\kappa_{k,n}+1}}e^{{\jmath}\frac{2\pi}{\lambda}d_{k,n}}, \forall k,n.
\end{align}
In contrast, the AoDs and path responses for the remaining $L_{k,n}-1$ NLoS paths change dynamically over time. Let $\mathbf{x}\triangleq\left[ x_1, \ldots, x_N\right] ^T\in \mathbb{R}^{N \times 1}$ denote all PA positions. Define the path response matrix as
\begin{equation}
	\mathbf{B}_k \triangleq \begin{bmatrix}
		\mathbf{b}_{k,1} & \mathbf{0}_{L_{k,1}} & \cdots & \mathbf{0}_{L_{k,1}} \\
		\mathbf{0}_{L_{k,2}} & \mathbf{b}_{k,2} & \cdots & \mathbf{0}_{L_{k,2}} \\
		\vdots & \vdots & \ddots & \vdots \\
		\mathbf{0}_{L_{k,N}} & \mathbf{0}_{L_{k,N}} & \cdots & \mathbf{b}_{k,N}
	\end{bmatrix} \in \mathbb{C}^{L_k \times N},
\end{equation}
where $L_k = \sum\nolimits_{n = 1}^N {{L_{k,n}}} $. By jointly considering in-waveguide propagation and free-space propagation, the statistical model for the channel from the PASS to user $k$ can be expressed as
\begin{subequations}
	\label{channel}
\begin{align}
	&{\mathbf{h}_k}\left( \mathbf{x} \right) = \mathbf{B}_k^H{\mathbf{f}_k}\left( \mathbf{x} \right) \in \mathbb{C}^{N \times 1},\forall k,\\
	&{\mathbf{f}_k}\left( \mathbf{x} \right) = {\left[ {{g_1}\left( {{x_1}} \right)\mathbf{f}_{k,1}^T\left( {{x_1}} \right), \ldots ,{g_N}\left( {{x_N}} \right)\mathbf{f}_{k,N}^T\left( {{x_N}} \right)} \right]^T}\in \mathbb{C}^{L_k \times 1}, \label{fk}\\
	&b_{k,n}^l \sim \mathcal{CN}\left( {0,\frac{{{\beta _0}d_{k,n}^{ - \alpha }}}{{\left( {{\kappa _{k,n}} + 1} \right)\left( {{L_{k,n}} - 1} \right)}}} \right),\forall 2 \le l \le L_{k,n},\forall k,n,\\
	&\theta _{k,n}^l \sim \mathcal{U}\left[ {0,\pi } \right],\forall 2 \le l \le L_{k,n},\forall k,n.
\end{align}
\end{subequations}
\subsection{Two-Timescale Antenna Positioning}
\begin{figure}[!t]
	\centering
	\includegraphics[width=1\linewidth]{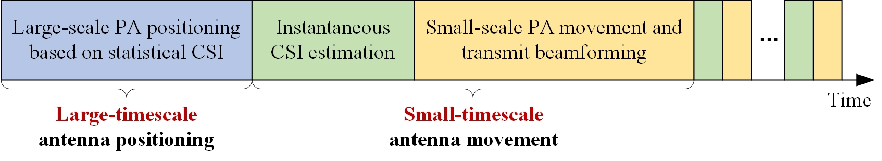}
	\caption{Illustration of the proposed two-timescale antenna positioning scheme.}
	\label{block}
\end{figure}
In Rician fading channels, the LoS component, which is determined by the PA positions and the user locations, varies slowly over time when the PAs and the users remain fixed or move only within small regions, and can thus be characterized statistically. In contrast, the NLoS component is random and generally varies more rapidly than the LoS component. Therefore, it is impractical to continuously acquire complete instantaneous CSI at arbitrary PA positions due to the prohibitively high complexity. Motivated by these observations, we propose a two-timescale antenna positioning scheme, as shown in Fig. \ref{block}. First, the large-timescale PA positions, denoted by $\mathbf{x}^\circ=\left[x^\circ_1, \ldots,x^\circ_N \right]^T \in \mathbb{R}^{N \times 1} $, are designed across the large-scale positioning region $\mathcal{X}_n^\circ$, $\forall n$, (large-scale PA positioning) to enhance long-term system performance based on statistical CSI, which is specified by the CSI of the LoS paths and given by
\begin{subequations}
	\label{sta_CSI}
\begin{align}
	&\bar{\mathbf{H}}\left( \mathbf{x}^\circ\right)=\left[ \bar{\mathbf{h}}_1\left(  \mathbf{x}^\circ\right), \ldots,\bar{\mathbf{h}}_K\left(  \mathbf{x}^\circ\right) \right] \in\mathbb{C}^{N \times K},\\
	&\bar{\mathbf{h}}_k(\mathbf{x}^\circ)
	=
	\left[
	\begin{array}{c}
		\sqrt{\frac{\beta_0d_{k,1}^{-\alpha}\kappa_{k,1}}{\kappa_{k,1}+1}}
		\rho_1 e^{\jmath 2\pi\left(\frac{1}{\lambda}d_{k,1}
			+\frac{1}{\lambda_{\mathrm{W}}}x_1^\circ\right)}
		\\
		\vdots
		\\
		\sqrt{\frac{\beta_0d_{k,N}^{-\alpha}\kappa_{k,N}}{\kappa_{k,N}+1}}
		\rho_N e^{\jmath 2\pi\left(\frac{1}{\lambda}d_{k,N}
			+\frac{1}{\lambda_{\mathrm{W}}}x_N^\circ\right)}
	\end{array}
	\right]
	\in\mathbb{C}^{N\times 1}.
\end{align}
\end{subequations}
After that, the instantaneous CSI is acquired for the fixed PA positions $\mathbf{x}^\circ$, rather than for arbitrary PA positions, thereby providing the basis for subsequent small-timescale PA position optimization\footnote{In general, small-scale PA positioning can be implemented using microelectromechanical systems (MEMS) with response times ranging from microseconds to milliseconds, or electronically reconfigurable elements with response times ranging from nanoseconds to microseconds \cite{MATutorial}. Accordingly, we assume that instantaneous channel measurement and reconstruction, together with small-scale PA positioning (including position optimization, PA movement, and settling), can be completed within the channel coherence time, which is a prerequisite for the applicability of the proposed small-timescale design. For channels that vary too rapidly to support small-scale PA positioning, the proposed two-timescale antenna positioning scheme in Fig. \ref{block} reduces to the conventional single-timescale design \cite{PASS14}, where the PA positions are optimized only on the large timescale.}. Specifically, similar to the channel estimation methods for movable antenna (MA) systems, the activated PA can be moved to a sparse set of pre-designed positions within its local small-scale movement region for channel measurements. Based on the channel measurement results, the channel response over the entire local region can be reconstructed by sparse signal processing methods, such as compressed sensing\footnote{In this paper, the channel estimation overheads are neglected to simplify the analysis and focus on the performance of two-timescale antenna positioning. Existing studies have developed efficient channel estimation schemes for MA-enabled communication systems under prescribed latency/energy constraints \cite{channel_est0}. The detailed design of practical channel estimation methods with reduced latency and energy consumption for the considered system is beyond the scope of this paper and is left for future investigation.} \cite{channel_est1,channel_est2}. Then, the small-timescale PA positions $\mathbf{x} =\left[x_1, \ldots,x_N \right]^T$ are optimized within the small-scale movement region $\mathcal{X}_n$, $\forall n$, based on the obtained instantaneous CSI to capture dynamic NLoS channel fluctuations in real time.
\subsection{Signal Model}
Denoting $\mathbf{H}\left(\mathbf{x} \right)=\left[\mathbf{h}_1\left(\mathbf{x} \right),\ldots,\mathbf{h}_K\left(\mathbf{x} \right) \right]\in \mathbb{C}^{N\times K}$ as the channel matrix from the PASS to all $K$ users, the received signal at the users can be expressed as
\begin{equation}
	\mathbf{y} = \mathbf{H}\left(\mathbf{x} \right)^H\mathbf{W}\mathbf{s}+\mathbf{n} ,
\end{equation}
where $\mathbf{W}=\left[\mathbf{w}_1,\ldots,\mathbf{w}_K \right]\in \mathbb{C}^{N\times K} $ is the transmit beamforming matrix at the PASS, $\mathbf{s}=\left[ s_1,\ldots,s_K\right]^T \in \mathbb{C}^{K\times 1}$ denotes the normalized information-bearing signals satisfying $\mathbb{E}\left(\mathbf{s}\mathbf{s}^H \right) =\mathbf{I}_K$, $\mathbf{n}=\left[ n_1,\ldots,n_K\right]^T \in \mathbb{C}^{K\times 1} \sim \mathcal{CN}\left(\mathbf{0}_K,\sigma^2 \mathbf{I}_K \right) $ represents the additive Gaussian noise at the users with average power $\sigma^2$. Then, the receive signal-to-interference-plus-noise ratio (SINR) of user $k$ can be written as
\begin{equation}\label{sinr}
	{\Gamma _k} = \frac{{{{\left| {{\mathbf{h}_k\left(\mathbf{x} \right)^H}\mathbf{w}_k} \right|}^2}}}{{\sum\limits_{k' = 1,k' \ne k}^K {{{\left| {{\mathbf{h}_k\left(\mathbf{x} \right)^H}\mathbf{w}_{k'}} \right|}^2}}  + {\sigma ^2}}}. 
\end{equation}
\subsection{Problem Formulation}
With \eqref{sinr}, the instantaneous rate for user $k$ is given by $R_k=\log_2{\left( 1+\Gamma _k\right) }$. Under the proposed two-timescale antenna positioning scheme, the objective is to maximize the ergodic achievable rate of $K$ users by jointly optimizing the two-timescale PA positions and the transmit beamforming matrix. Accordingly, the optimization problem can be formulated as\footnote{In this paper, perfect statistical CSI and perfect instantaneous CSI within the small-scale movement region are assumed for large-timescale and small-timescale PA position optimization, respectively, to characterize the theoretical performance limit of the proposed two-timescale antenna positioning scheme. Nevertheless, the impact of imperfect instantaneous CSI on PA position optimization will be evaluated via simulations in Section \ref{mu_section}.}
\begin{subequations}
	\label{max1}
	\begin{align}
		& \mathop {\mathrm{max} }\limits_{\mathbf{x}^\circ} \: \mathbb{E}\left(\mathop {\max }\limits_{\mathbf{x},\mathbf{W}}  \mathop {\min }\limits_{1 \le k \le K} {R _k}\right) \label{obj1} \\
		&\mathrm{s.t.}\quad \left\|\mathbf{W}\right\|_F^2 \le P,\label{c1} \\
		&\hspace{2.3em} x_n^\circ \in \mathcal{X}_n^\circ,  \forall n,\label{c2} \\
		&\hspace{2.3em} x_n \in \mathcal{X}_n,  \forall n,\label{c3}
	\end{align}
\end{subequations}
where $\mathbb{E}\left(\cdot\right) $ in \eqref{obj1} represents the expectation over random NLoS channel realizations. Constraint \eqref{c1} indicates that the transmit power of the PASS cannot exceed its power budget $P$. Constraints \eqref{c2} and \eqref{c3} respectively confine the large-timescale and small-timescale PA positions within finite ranges. Problem \eqref{max1} is challenging to solve because it is a non-convex optimization problem with coupled variables. Thus, we first address the special case of the single-user scenario, and then extend the results to the general multi-user case.
\section{Single-User System}\label{3}
This section considers the special single-user setup, i.e., $K=1$, to draw crucial insights. For brevity, the user index $k$ is omitted. In this case, there is no inter-user interference, and thus the maximum ratio transmission (MRT) method is optimal, which is given by 
\begin{equation}\label{MRT}
	\mathbf{w}\left(\mathbf{x} \right)=\sqrt{P}\frac{\mathbf{h}\left(\mathbf{x} \right)}{\left\|\mathbf{h}\left(\mathbf{x} \right)\right\|_2}.
\end{equation}
Then, the user rate is simplified as $R=\log_2{\left( 1+{P}\left\|\mathbf{h}\left(\mathbf{x} \right)\right\|_2^2/{\sigma^2}\right) }$. Therefore, problem \eqref{max1} can be reformulated as
\begin{equation}\label{max2}
	\mathop {\mathrm{max} }\limits_{\mathbf{x}^\circ} \: \mathbb{E}\left(\mathop {\max }\limits_{\mathbf{x}} \: \left\|\mathbf{h}\left(\mathbf{x} \right)\right\|_2^2\right) \quad\mathrm{s.t.} \: \eqref{c2},\eqref{c3}.
\end{equation}
It is worth noting that the PA can reduce path loss through large-scale positioning. To maximize the ergodic channel gain, the PAs need to be positioned as close to the user as possible. Thus, the optimal large-timescale PA positions are given by
\begin{equation}\label{large_optimal}
	x_n^\circ = \min\left(\max\left(X,\frac{D_n}{2} \right),S_\mathrm{x}-\frac{D_n}{2}  \right) ,\forall n,
\end{equation}
which implies that the PAs are located on a straight line directly above the user and parallel to the $y$-axis. Thereafter, the position of each PA is refined within the range $\mathcal{X}_n$ to maximize the instantaneous channel gain.
\subsection{Performance Analysis}\label{pf}
\begin{figure*}[!b]
	\textsc{\centering
		\hrulefill
		\begin{align}\label{channel_gain}
			G_n\left( x_n\right) &=	 {{{\left( {{g_n}\left(x_n \right) \mathbf{b}_n^H{\mathbf{f}_n\left(x_n \right)}} \right)}^*}} \left( {{g_n\left(x_n \right)}\mathbf{b}_n^H{\mathbf{f}_n\left(x_n \right)}} \right) \nonumber\\
			&= {\left( {\sum\limits_{l = 1}^{{L_n}} {b_n^l{\rho _n}{e^{ - \jmath \left( {\frac{{2\pi }}{\lambda }\left( {{x_n} - x_n^\circ} \right)\cos\theta _n^l + \frac{{2\pi }}{{{\lambda _\mathrm{W}}}}{x_n}} \right)}}} } \right)\left( {\sum\limits_{l = 1}^{{L_n}} {{{\left( {b_n^l} \right)}^*}{\rho _n}{e^{\jmath \left( {\frac{{2\pi }}{\lambda }\left( {{x_n} - x_n^\circ} \right)\cos\theta _n^l + \frac{{2\pi }}{{{\lambda _\mathrm{W}}}}{x_n}} \right)}}} } \right)}\nonumber\\
			&=  {\rho _n^2\sum\limits_{l = 1}^{{L_n}} {{{\left| {b_n^l} \right|}^2}} }  +  {\rho _n^2\sum\limits_{i = 1}^{{L_n} - 1} {\sum\limits_{j = i + 1}^{{L_n}} {2\left| {b_n^i{{\left( {b_n^j} \right)}^*}} \right|\cos \left( {\frac{{2\pi }}{\lambda }\left( {{x_n} - x_n^\circ} \right)\left( {\cos\theta _n^j - \cos\theta _n^i} \right) + \angle \left( {b_n^i{{\left( {b_n^j} \right)}^*}} \right)} \right)} } } \nonumber\\
			&\mathop  = \limits^{\left( \mathrm{a} \right)} A_n  +  {\rho _n^2\sum\limits_{i = 1}^{{L_n} - 1} {\sum\limits_{j = i + 1}^{{L_n}} {2\left| b_n^{i,j} \right|\cos \left( {\frac{{2\pi }}{\lambda } \tilde{x}_n  \vartheta _n^{i,j}  + \angle b_n^{i,j} } \right)} } }.
		\end{align}
	}
\end{figure*}
In the following, the upper bound on the channel gain and the conditions for achieving this bound are analyzed for the given large-timescale PA positions under different numbers of channel paths. For the convenience of subsequent analysis, the channel gain is rewritten as $\left\|\mathbf{h}\left(\mathbf{x}\right)\right\|_2^2=\sum\nolimits_{n=1}^N G_n\left(x_n \right) $, where $G_n\left(x_n \right)$ represents the channel gain between PA $n$ and the user, which is derived in \eqref{channel_gain} at the bottom of the next page. In \eqref{channel_gain}, the equality ($\mathrm{a}$) holds by defining
\begin{subequations}
	\begin{align}
		A_n&\triangleq{\rho _n^2\sum\limits_{l = 1}^{L_n} {{{\left| {b_n^l} \right|}^2}} },\forall n,\\
		b_n^{i,j}&\triangleq{b_n^{i}{{\left( {b_n^j} \right)}^*}},\forall n,i,j,\\
		\tilde {x}_n&\triangleq{{x_n} - x_n^\circ} \in \tilde{\mathcal{X}}_n,\forall n,\\
		\tilde{\mathcal{X}}_n&\triangleq \left[-D _{n}/2,D _{n}/2\right],\forall n,\\
		\vartheta _n^l &\triangleq \cos \theta _n^l,\forall n,l,\\
		\vartheta _n^{i,j}&\triangleq{\vartheta _n^j - \vartheta _n^i},\forall n,i,j.
	\end{align}
\end{subequations}
Since the movement of each PA independently affects only its own channel to the user, the upper bound can be achieved by maximizing $G_n\left(x_n \right)$ for each PA.
\subsubsection{One-Path Case}
For the one-path case, i.e., $L_n = 1$, $\forall n$, the channel gain \eqref{channel_gain} reduces to
\begin{equation}
	{G^{L_n = 1}_n}  =  {\rho _n^2{{\left| {b_n^1} \right|}^2}}  =  {\rho _n^2 \beta_0d_{n}^{-\alpha}} ,
\end{equation}
which indicates that the small-timescale PA position refinement does not improve the channel gain. Therefore, in the one-path scenario, the optimal small-timescale PA positions are exactly given by $x_n^\star = x_n^\circ$, $\forall n$.
\subsubsection{Two-Path Case}
For the two-path case, i.e., $L_n = 2$, $\forall n$, the channel gain \eqref{channel_gain} can be reformulated as
\begin{equation}\label{channel_gain2}
	{G^{L_n = 2}_n}\left( \tilde{x}_n\right)= A_n +  {{2\rho_n^2\left| b_n^{1,2} \right|\cos \left( {\frac{{2\pi }}{\lambda } \tilde {x}_n  \vartheta _n^{1,2}  + \angle b_n^{1,2}} \right)}},
\end{equation}
which is a periodic function with period ${\lambda }/{{\left| {\vartheta _n^{1,2}} \right|}}$. To analyze the variation of ${G^{L_n = 2}_n}\left( \tilde{x}_n\right)$ within $\tilde{\mathcal{X}}_n$, the first-order derivative of which with respect to $\tilde {x}_n$ is derived as
\begin{equation}
	\frac{\mathrm{d}{G^{L_n = 2}_n}\left( \tilde{x}_n\right)}{\mathrm{d}\tilde {x}_n}= - \frac{{4\pi\rho_n^2 \left| {b_n^{1,2}} \right|\vartheta _n^{1,2}}}{\lambda }\sin \left( {\frac{{2\pi }}{\lambda } \tilde {x}_n  \vartheta _n^{1,2}  + \angle b_n^{1,2}} \right). 
\end{equation}
The sufficient condition for ${\mathrm{d}{G^{L_n = 2}_n}\left( \tilde{x}_n\right)}/{\mathrm{d}\tilde {x}_n}\le 0$ is given by
\begin{equation}
	\left\{
	\begin{array}{l}
		\frac{\lambda}{\vartheta^{1,2}_n} \left( z_1 - \frac{\angle {b_n^{1,2}}}{2\pi} \right) \le \tilde {x}_n \le \frac{\lambda}{\vartheta^{1,2}_n} \left( z_1  -\frac{\angle {b_n^{1,2}}}{2\pi} + \frac{1}{2} \right),  \vartheta_n^1 \le \vartheta_n^2, \\
		\frac{\lambda}{\vartheta^{1,2}_n} \left( z_1 - \frac{\angle {b_n^{1,2}}}{2\pi} \right) \le \tilde {x}_n \le \frac{\lambda}{\vartheta^{1,2}_n} \left( z_1   -\frac{\angle {b_n^{1,2}}}{2\pi} - \frac{1}{2}\right),  \vartheta_n^1 > \vartheta_n^2,
	\end{array}
	\right.
\end{equation}
where $ z_1 \in \mathbb{Z}$. Define $x_n^\mathrm{e}$ as a maximum point of ${G^{L_n = 2}_n}\left( \tilde{x}_n\right)$ over $\tilde{\mathcal{X}}_n$. Thus, the channel gain ${G^{L_n = 2}_n}\left( \tilde{x}_n\right)$ decreases as $\tilde {x}_n$ increases within $\left[ x_n^\mathrm{e},D_n/2\right] $ if $\exists z_1 \in \mathbb{Z}$ such that
\begin{equation}\label{ine1}
	\left\{
	\begin{array}{l}
		\frac{{{D_n}\vartheta _n^{1,2}}}{{2\lambda }} + \frac{{\angle b_n^{1,2}}}{{2\pi }} - \frac{1}{2} \le {z_1} \le \frac{{x_n^\mathrm{e}\vartheta _n^{1,2}}}{\lambda } + \frac{{\angle b_n^{1,2}}}{{2\pi }},  \vartheta_n^1 \le \vartheta_n^2, \\ 
		\frac{{x_n^\mathrm{e}\vartheta _n^{1,2}}}{\lambda } + \frac{{\angle b_n^{1,2}}}{{2\pi }} \le {z_1} \le \frac{{{D_n}\vartheta _n^{1,2}}}{{2\lambda }} + \frac{{\angle b_n^{1,2}}}{{2\pi }} + \frac{1}{2},  \vartheta_n^1 > \vartheta_n^2,
	\end{array}
	\right.
\end{equation}
Moreover, to ensure that inequality \eqref{ine1} holds, $x_n^\mathrm{e}$ and $D_n$ need to satisfy
\begin{equation}\label{range1}
	\max \left( {\frac{{{D_n}}}{2} - \frac{\lambda }{{2\left| {\vartheta _n^{1,2}} \right|}},-\frac{{{D_n}}}{2}} \right) \le x_n^\mathrm{e} \le \frac{{{D_n}}}{2}.
\end{equation}
Similarly, the sufficient condition for ${\mathrm{d}{G^{L_n = 2}_n}\left( \tilde{x}_n\right)}/{\mathrm{d}\tilde {x}_n}\ge 0$ is given by
\begin{equation}
	\left\{
	\begin{array}{l}
		\frac{\lambda}{\vartheta^{1,2}_n} \left( z_2 - \frac{\angle {b_n^{1,2}}}{2\pi}-\frac{1}{2} \right) \le \tilde {x}_n \le \frac{\lambda}{\vartheta^{1,2}_n} \left( z_2 -\frac{\angle {b_n^{1,2}}}{2\pi} \right),  \vartheta_n^1 \le \vartheta_n^2, \\
		\frac{\lambda}{\vartheta^{1,2}_n} \left( z_2 - \frac{\angle {b_n^{1,2}}}{2\pi}+\frac{1}{2} \right)\le \tilde {x}_n \le \frac{\lambda}{\vartheta^{1,2}_n} \left( z_2 -\frac{\angle {b_n^{1,2}}}{2\pi} \right),  \vartheta_n^1 > \vartheta_n^2,
	\end{array}
	\right.
\end{equation}
where $ z_2 \in \mathbb{Z}$. The channel gain ${G^{L_n = 2}_n}\left( \tilde{x}_n\right)$ increases with $\tilde {x}_n$ within $\left[ -D_n/2,x_n^\mathrm{e}\right]   $ if $\exists z_2 \in \mathbb{Z}$ such that 
\begin{equation} \label{ine2}
	\left\{
	\begin{array}{l}
		\frac{{x_n^\mathrm{e}\vartheta _n^{1,2}}}{\lambda } + \frac{{\angle b_n^{1,2}}}{{2\pi }} \le {z_2} \le  - \frac{{{D_n}\vartheta _n^{1,2}}}{{2\lambda }} +\frac{{\angle b_n^{1,2}}}{{2\pi }} + \frac{1}{2},  \vartheta_n^1 \le \vartheta_n^2, \\ 
		- \frac{{{D_n}\vartheta _n^{1,2}}}{{2\lambda }} +\frac{{\angle b_n^{1,2}}}{{2\pi }} - \frac{1}{2} \le {z_2} \le \frac{{x_n^\mathrm{e}\vartheta _n^{1,2}}}{\lambda } + \frac{{\angle b_n^{1,2}}}{{2\pi }},  \vartheta_n^1 > \vartheta_n^2,
	\end{array}
	\right.
\end{equation}
where
\begin{equation}\label{range2}
	0 \le x_n^\mathrm{e} \le \min \left( {-\frac{{{D_n}}}{2}+\frac{\lambda }{{2\left| {\vartheta _n^{1,2}} \right|}},\frac{{{D_n}}}{2}} \right).
\end{equation}
With \eqref{range1} and \eqref{range2}, $D_n$ and ${\vartheta _n^{1,2}}$ need to satisfy
\begin{equation}
	{D_n} \le \frac{\lambda }{{\left| {\vartheta _n^{1,2}} \right|}},
\end{equation}
which indicates that the length $D_n$ of the range for small-timescale PA position refinement is no greater than the period ${\lambda }/{{\left| {\vartheta _n^{1,2}} \right|}}$ of the cosine function ${G^{L_n = 2}_n}\left( \tilde{x}_n\right)$ in \eqref{channel_gain2}. In this case, if $\exists z_1,z_2 \in \mathbb{Z}$ that satisfy inequalities \eqref{ine1} and \eqref{ine2}, the optimal small-timescale PA position for maximizing the channel gain \eqref{channel_gain2} is $x_n^\star=x_n^\mathrm{e}+x_n^\circ$, $\forall n$. On the other hand, when ${D_n} > {\lambda }/{{\left| {\vartheta _n^{1,2}} \right|}}$, there may be multiple optimal small-timescale PA positions, which are given by
\begin{equation}
	x_n^\star=\frac{\lambda }{{\vartheta _n^{1,2}}}\left( {{z_3} - \frac{{\angle b_n^{1,2}}}{{2\pi }}} \right)+x_n^\circ, \forall n,
\end{equation}
where $z_3 \in \mathbb{Z}$ such that $x_n^\star \in \mathcal{X}_n$, $\forall n$.
\subsubsection{Multiple-Path Case}
For the multiple-path case, i.e., $L_n > 2$, $\forall n$, the channel gain \eqref{channel_gain} can be expressed as
\begin{equation}\label{channel_gain3}
	{G^{L_n > 2}_n}\left( \tilde{x}_n\right) =  {\rho _n^2{{\left| {\sum\limits_{l = 1}^{{L_n}} {{{\left( {b_n^l} \right)}^*}{e^{\jmath \frac{{2\pi }}{\lambda }\vartheta _n^l{{\tilde x}_n}}}} } \right|}^2}} .
\end{equation}
It is difficult to explicitly represent the period of ${G^{L_n > 2}_n}\left( \tilde{x}_n\right)$ due to the random AoDs. Nevertheless, the approximation of this period can be derived by assuming that the virtual AoDs are quantized. Let $\tau_n$ denote the period of ${G^{L_n > 2}_n}\left( \tilde{x}_n\right)$. Then, we have
\begin{align}
	&{{{\left| {\sum\limits_{l = 1}^{{L_n}} {{{\left( {b_n^l} \right)}^*}{e^{\jmath \frac{{2\pi }}{\lambda }\vartheta _n^l{{\tilde x}_n}}}} } \right|}^2}}={{{\left| {\sum\limits_{l = 1}^{{L_n}} {{{\left( {b_n^l} \right)}^*}{e^{\jmath \frac{{2\pi }}{\lambda }\vartheta _n^l{\left( {\tilde x}_n+\tau_n\right) }}}} } \right|}^2}}, \nonumber\\
	\Leftrightarrow & \sum\limits_{i = 1}^{{L_n}} {\sum\limits_{j = 1}^{{L_n}} {b_n^{i,j}{e^{\jmath \frac{{2\pi }}{\lambda }\vartheta _n^{i,j}{{\tilde x}_n}}}} }  = \sum\limits_{i = 1}^{{L_n}} {\sum\limits_{j = 1}^{{L_n}} {b_n^{i,j}{e^{\jmath\frac{{2\pi }}{\lambda }\vartheta _n^{i,j}\left( {{{\tilde x}_n} + {\tau _n}} \right)}}} } ,\nonumber\\
	\Leftrightarrow & \sum\limits_{i = 1}^{{L_n}} {\sum\limits_{j = 1,j \ne i}^{{L_n}} {b_n^{i,j}{e^{\jmath \frac{{2\pi }}{\lambda }\vartheta _n^{i,j}{{\tilde x}_n}}}\left( {1 - {e^{\jmath \frac{{2\pi }}{\lambda }\vartheta _n^{i,j}{\tau _n}}}} \right)} }  = 0,\nonumber\\
	\Leftrightarrow & {1 - {e^{\jmath \frac{{2\pi }}{\lambda }\vartheta _n^{i,j}{\tau _n}}}} =0, 1 \le i,j \le L_n,\nonumber\\
	\Leftrightarrow & \frac{\vartheta _n^{i,j}{\tau _n}}{\lambda} \in \mathbb{Z}, 1 \le i,j \le L_n,
\end{align}
which indicates that period $\tau _n$ is the minimum real number that ensures ${\vartheta _n^{i,j}{\tau _n}}/{\lambda} \in \mathbb{Z}$, $\forall i,j$. The virtual AoDs $\{\vartheta _n^l\}_{l=1}^{L_n}$ are quantized with a resolution of $\eta_n^0$, i.e., ${\vartheta _n^l} \in {\{ { - 1 + {{\left( 2\eta_n - 1\right)  }}/{\eta_n^0}} \}_{\eta_n=1}^{\eta_n^0}}$, and sorted in a non-decreasing order, i.e., ${\vartheta_n^1} \le {\vartheta_n^2} \le  \ldots  \le {\vartheta_n^{L_n}}$. Denote $\vartheta_n^l=- 1 + {{\left( 2\eta_n^l - 1\right)  }}/{\eta_n^0}$, indicating that $\vartheta_n^l$ corresponds to the $\eta_n^l$-th element in the quantized set of virtual AoDs. Then, the difference of two adjacent virtual AoDs can be obtained as ${\vartheta ^{l + 1}_n} - {\vartheta^l_n} ={2}\left( {{\eta^{l + 1}_n} - {\eta_n^l}} \right)/{\eta_n^0}  = {{2}}\Delta _n^l/{\eta_n^0}$, $1\le l \le L_n-1$. To guarantee ${\vartheta _n^{i,j}{\tau _n}}/{\lambda} \in \mathbb{Z}$, $\forall i,j$, the minimum period of ${G^{L_n > 2}_n}\left( \tilde{x}_n\right)$ should be given by 
\begin{equation}
	\tau _n = \frac{\eta_n^0\lambda}{2\mu_n},
\end{equation}
where $\mu_n$ is the maximal common factor for $\{{\Delta _n^l}\}_{l=1}^{L_n-1}$. Thus, the maximum value of ${G^{L_n > 2}_n}\left( \tilde{x}_n\right)$ can always be achieved within any region where $\tilde{x}_n$ spans at least one period $\tau _n$. Under the assumption of quantized virtual AoDs, the optimal small-timescale PA position must lie in the interval
\begin{equation}
	x_n^\star\! \in\! \left[  \max\!\left(\! x_n^\circ \! - \! \frac{D_n}{2},  x_n^\circ  \! -  \!\frac{\tau_n}{2}  \! \right)  , \min \!\left(  \! x_n^\circ \! + \! \frac{D_n}{2}  ,  x_n^\circ \! + \! \frac{\tau_n}{2} \! \right)  \right],\forall n.
\end{equation}
\subsection{Optimization Algorithm}
For the single-user system, the optimal large-timescale PA position is given by \eqref{large_optimal}. Subsequently, a parallel search algorithm is proposed to optimize the small-timescale PA position. For any PA $n$, its small-scale movement region $\mathcal{X}_n$ is discretized into $Q_n$ points. The set of candidate positions is given by
\begin{equation}
	\mathcal{S}_n=\left\{x_n^\circ-\frac{D_n}{2}+\frac{D_n }{Q_n-1}q_n\middle|
	q_n \in \left\{0,\ldots,Q_n-1\right\}\right\},
\end{equation}
where $q_n $ denotes the position index. Since the movement of each PA independently affects only the channel gain \eqref{channel_gain}, the small-timescale positions of all PAs can be optimized in parallel by 
\begin{equation}\label{xn_small}
	x_n = \arg \mathop {\max }\limits_{x_n \in \mathcal{S}_n}  G_n\left(x_n \right) ,
\end{equation}
instead of being optimized in an alternating manner.
\subsection{Overall Algorithm}
\begin{algorithm}[!t]
	\caption{Proposed algorithm for solving problem \eqref{max2}}
	\label{alg1}
	\renewcommand{\algorithmicrequire}{\textbf{Input:}}
	\renewcommand{\algorithmicensure}{\textbf{Output:}}
	\begin{algorithmic}[1]
		\REQUIRE $N$, $\lambda$, $h$, $S_\mathrm{x}$, $n_\mathrm{e}$, $\sigma^2$, $P$, $\alpha$, $\beta_0$, $\mathbf{u}$, $\{y_n\}$, $\{\rho_n\}$, $\{L_{n}\}$, $\{\theta_{n}^l\}$, $\{b_{n}^l\}$, $\{D_n\}$, $\{d_{n}\}$, $\{\kappa_{n}\}$, $\{Q_n\}$.
		\ENSURE $\mathbf{x}$, $\mathbf{w}$.
		\STATE Obtain the large-timescale position of each PA via \eqref{large_optimal}.
		\STATE Optimize the small-timescale positions of all PAs in parallel via \eqref{xn_small}.
		\STATE Obtain the transmit beamforming vector via \eqref{MRT}.
		\RETURN $\mathbf{x}$, $\mathbf{w}$.
	\end{algorithmic}
\end{algorithm}
The overall algorithm for solving problem \eqref{max2} is detailed in Algorithm \ref{alg1}. In line 1, the optimal large-timescale PA positions are obtained via \eqref{large_optimal}, with the complexity being $\mathcal{O}( N)$. In line 2, the small-timescale positions of all PAs are searched in parallel via \eqref{xn_small}. Since the complexity for calculating the channel gain $G_n\left(x_n \right)$ using \eqref{channel_gain3} is $\mathcal{O}( L_n) $, the overall complexity of line 2 is $\mathcal{O}( \sum\nolimits_{n=1}^N Q_n L_n) $. In line 3, the transmit beamforming vector is obtained via \eqref{MRT}. Thus, the total computational complexity of Algorithm \ref{alg1} is $\mathcal{O}( N+\sum\nolimits_{n=1}^N Q_n L_n) $.
\section{Multi-User System}\label{4}
This section considers the general multi-user scenario, i.e., $K>1$. An upper bound on the minimum achievable rate among all users is first derived for fixed large-timescale PA positions. Then, the optimization algorithm for the single-user case is extended to obtain an efficient solution to problem \eqref{max1}.
\subsection{Performance Analysis}
For the fixed large-timescale PA positions, if the small-timescale PA position refinement nulls the inter-user interference and the received signal power for each user reaches its maximum value, the receive SINR \eqref{sinr} attains an upper bound. Specifically, the maximum received signal power for each user can be achieved by MRT, i.e., $\mathbf{w}_k\left(\mathbf{x}\right) =\sqrt{p_k}\frac{\mathbf{h}_k\left(\mathbf{x}\right) }{\left\|\mathbf{h}_k\left(\mathbf{x}\right)\right\|_2}$, where $p_k$ denotes the transmit power allocated to user $k$. Therefore, the receive SINR \eqref{sinr} is upper-bounded by
\begin{equation}\label{ub1}
	\Gamma_k\left(\mathbf{x} \right) \mathop  \le \limits^{\left( {{\mathrm{b}_1}} \right)}  \frac{{{{\left| {{\mathbf{h}_k}\left( \mathbf{x}\right)^H\mathbf{w}_k\left( \mathbf{x}\right)  } \right|}^2}}}{{ {\sigma ^2}}} \mathop  \le \limits^{\left( {{\mathrm{b}_2}} \right)} \frac{p_k}{\sigma^2}\left\|\mathbf{h}_k\left(\mathbf{x} \right)\right\|_2^2,
\end{equation}
where the equality holds in ($\mathrm{b}_1$) if the channel vectors of any two users are orthogonal, i.e., 
\begin{equation}\label{orthogonal}
	\mathbf{h}_k\left( {\mathbf{x}} \right)^H{\mathbf{h}_{k'}}\left( {\mathbf{x}} \right) = 0, \forall 1 \le k \ne k' \le K,
\end{equation}
and the equality holds in ($\mathrm{b}_2$) if the PASS employs MRT to maximize the received signal power of the users. In addition, the channel gain $\left\|\mathbf{h}_k\left(\mathbf{x} \right)\right\|_2^2$ in \eqref{ub1} is upper-bounded by
\begin{equation}\label{gain}
	\left\| {{\mathbf{h}_k}\left( \mathbf{x} \right)} \right\|_2^2 = \sum\limits_{n = 1}^N {{{\left| {\mathbf{b}_{k,n}^H{\mathbf{f}_{k,n}}\left( {{x_n}} \right){g_n}\left( {{x_n}} \right)} \right|}^2}} \mathop  \le \limits^{\left( \mathrm{c} \right)} \sum\limits_{n = 1}^N {\left\| {{\rho _n}{\mathbf{b}_{k,n}}} \right\|_1^2},  
\end{equation}
where the equality holds in ($\mathrm{c}$) if ${\mathbf{f}_{k,n}}\left( {{x_n}} \right){g_n}\left( {{x_n}} \right) = {\rho _n}{e^{\jmath \angle {\mathbf{b}_{k,n}}}}$, $\forall n$. Based on \eqref{ub1} and \eqref{gain}, an upper bound on the receive SINR \eqref{sinr} can be obtained as
\begin{equation}
	\Gamma_k^\mathrm{ub} =\frac{p_k}{\sigma^2}\sum\limits_{n = 1}^N {\left\| {{\rho _n}{\mathbf{b}_{k,n}}} \right\|_1^2}.
\end{equation}

To maximize the minimum achievable rate among all $K$ users for any given instantaneous channel realization, the transmit power should be allocated such that all users have the same achievable rate. Otherwise, the transmit power of the users with higher achievable rates can always be reduced and reallocated to those with lower achievable rates to improve the max-min rate. Thus, the optimal power allocation for maximizing the minimum instantaneous rate can be obtained by solving 
\begin{equation}
	\left\{
	\begin{array}{l}
		{{{\log }_2}\left( {1 + \Gamma_k^\mathrm{ub}} \right) 
			= R_\mathrm{ub}} ,\\
		{\sum\limits_{k = 1}^K {{p_k}}  = P},
	\end{array}
	\right.
\end{equation}
where $R_\mathrm{ub}$ denotes an upper bound on the minimum instantaneous rate, which can be derived as
\begin{equation}\label{R_tup}
	R_\mathrm{ub}={\log _2}\left( {1 + \frac{P}{{{\sigma ^2}\sum\limits_{k = 1}^K {{{\left( {\sum\limits_{n = 1}^N {\left\| {{\rho _n}{\mathbf{b}_{k,n}}} \right\|_1^2} } \right)}^{ - 1}}} }}} \right).
\end{equation}
For the special case where only a single path exists from each PA to each user, the optimal small-timescale PA positions that achieve the upper bound \eqref{R_tup} are given by
\begin{equation}\label{pnset}
	\mathbf{p}_n  \in \left\{ \mathbf{p} \,\middle|\,
	\begin{aligned}
		& \| \mathbf{p} - \mathbf{u}_{k'} \|_2 - \| \mathbf{p} - \mathbf{u}_k \|_2 
		= \lambda \left( z_{k,k',n} + \phi_{k,k',n} \right)\\ 
		& \forall k \ne k', \forall z_{k,k',n} \in \mathbb{Z}, 
		\forall 0 \le \phi_{k,k',n} < 1
	\end{aligned}
	\right\},
\end{equation}
with ${\sum\nolimits_{n = 1}^N {{\beta _0} \rho_n^2\sqrt {\left( d_{k,n}d_{k',n}\right) ^{ - \alpha }} {e^{\jmath 2\pi {\phi _{k,k',n}}}}}  = 0 }$, and the proof is provided in Appendix \ref{appendixA}.

Nevertheless, under certain channel conditions, the set \eqref{pnset} may be empty, in which case the upper bound \eqref{R_tup} cannot be achieved. Moreover, in the general case where multiple channel paths exist from the PAs to the users, it is difficult to derive an explicit solution for the PA positions. Therefore, in the next subsection, we develop efficient algorithms to optimize the PA positions at both the large and small timescales.
\subsection{Optimization Algorithm}
Thanks to the large array size in PASS, the channel correlation between users can be significantly mitigated through PA position optimization. Therefore, a low-complexity zero-forcing (ZF) method is employed to design the transmit beamforming matrix\footnote{Although ZF beamforming is not optimal, the numerical results in Section \ref{5} will demonstrate that the resulting rate performance approaches its upper bound. In addition, ZF beamforming has a closed-form expression, which reduces the computational complexity of PA position optimization. Hence, the ZF-based design achieves a good trade-off between optimality and complexity.}, which is given by
\begin{equation}\label{beamforming}
	\mathbf{W} \left( {\mathbf{x}} \right) = \sqrt P \frac{{\mathbf{H}\left( {\mathbf{x}} \right){{\left( {{\mathbf{H}}\left( {\mathbf{x}} \right)^H\mathbf{H}\left( {\mathbf{x}} \right)} \right)}^{ - 1}}}}{{{{\left\| {\mathbf{H}\left( {\mathbf{x}} \right){{\left( {{\mathbf{H}}\left( {\mathbf{x}} \right)^H\mathbf{H}\left( {\mathbf{x}} \right)} \right)}^{ - 1}}} \right\|}_F}}} \in \mathbb{C}^{N \times K},
\end{equation}
which yields the same achievable rate for all $K$ users as
\begin{equation}\label{R_small}
	R\left( {\mathbf{x}} \right)   = \log_2\left(1+ \frac{P}{{\mathrm{tr}\left( {{{\left( {{\mathbf{H}}\left( {\mathbf{x}} \right)^H\mathbf{H}\left( {\mathbf{x}} \right)} \right)}^{ - 1}}} \right){\sigma ^2}}}\right) .
\end{equation}
Then, the PA position optimization in problem \eqref{max1} is simplified as 
\begin{equation}\label{max3}
	\mathop {\mathrm{max} }\limits_{\mathbf{x}^\circ} \: \mathbb{E}\left(\mathop {\max }\limits_{\mathbf{x}} \: R\left( {\mathbf{x}} \right)\right) \quad\mathrm{s.t.} \: \eqref{c2},\eqref{c3}.
\end{equation}
Next, an element-wise search algorithm is proposed for large-timescale PA position optimization, while a projected gradient ascent algorithm is developed for small-timescale PA position refinement.
\subsubsection{Large-Timescale PA Position Optimization}
A conventional approach to handling the intractable expectation in problem \eqref{max3} is to apply the Monte Carlo approximation, which is given by
\begin{equation}\label{MC}
	\mathbb{E}\left( \max_{\mathbf{x}} \: R\left( {\mathbf{x}} \right) \right)
	\approx 
	\frac{1}{M} \sum_{m=1}^{M} 
	\max_{\mathbf{x}} \: R^m \left( {\mathbf{x}} \right),
\end{equation}
where $M$ is the total number of Monte Carlo simulations and $R^m\left( {\mathbf{x}} \right)$ denotes the achievable rate of the users under the $m$-th ($1 \le m \le M$) independent channel realization based on the known statistical CSI. The Monte Carlo approximated expectation in \eqref{MC} relies on a sufficiently large number of random channel realizations, which entails a high computational complexity. To achieve lower complexity, we directly approximate the expectation as
\begin{equation}\label{dire}
	\mathbb{E}\left( \max_{\mathbf{x}} \: R\left( {\mathbf{x}} \right) \right)
	\approx 
	 \bar{R}\left( {\mathbf{x}^\circ} \right),
\end{equation}
where $\bar{R}\left( {\mathbf{x}^\circ} \right)$ is determined based on the statistical CSI in \eqref{sta_CSI} as
\begin{equation}\label{Rk_sta}
	\bar{R}\left( {\mathbf{x}^\circ} \right)   = \log_2\left(1+ \frac{P}{{\mathrm{tr}\left( {{{\left( {{\bar{\mathbf{H}}}\left( {\mathbf{x}^\circ} \right)^H\bar{\mathbf{H}}\left( {\mathbf{x}^\circ} \right)} \right)}^{ - 1}}} \right){\sigma ^2}}}\right).
\end{equation}
Then, the large-timescale PA position optimization problem is given by 
\begin{equation}\label{max4}
	\mathop {\mathrm{min} }\limits_{\mathbf{x}^\circ} \:\mathrm{tr}\left( {{{\left( {{\bar{\mathbf{H}}}\left( {\mathbf{x}^\circ} \right)^H\bar{\mathbf{H}}\left( {\mathbf{x}^\circ} \right)} \right)}^{ - 1}}} \right) \quad\mathrm{s.t.} \: \eqref{c2}.
\end{equation}

To conduct the element-wise search for solving problem \eqref{max4}, the large-scale positioning region $\mathcal{X}_n^\circ$ of each PA is discretized into $\tilde{Q}_n$ points. Then, the set of candidate PA positions for PA $n$ is defined as
\begin{equation}
	\tilde{\mathcal{S}}_n=\left\{\frac{D_n}{2}+\frac{S_\mathrm{x}-D_n }{\tilde{Q}_n-1}\tilde{q}_n\middle|
	\tilde{q}_n \in \left\{0,\ldots,\tilde{Q}_n-1\right\}\right\},
\end{equation}
where $\tilde{q}_n$ denotes the position index. To alleviate the computational burden, these discretized positions are sparsely spaced. It is worth noting that the change in the position of PA $n$ affects only the elements of the $n$-th row of the matrix $\bar{\mathbf{H}}\left(\mathbf{x}^\circ \right) $. Thus, it is unnecessary to calculate the matrix inverse in \eqref{Rk_sta} for each candidate position of PA $n$, which incurs high computational complexity. Alternatively, the Sherman-Morrison lemma \cite{PASS_comm3} can be used for the efficient calculation of the matrix inverse. Let $\tilde{\mathbf{h}}_n\left( x_n^\circ\right) \in \mathbb{C}^{K \times 1}$ denote the $n$-th column of $\bar{\mathbf{H}}\left(\mathbf{x}^\circ \right)^H $. The matrix inverse in problem \eqref{max4} can be reformulated as 
\begin{align}\label{Sherman}
	&\left( \bar{\mathbf{H}}\left( \mathbf{x}^\circ\right) ^H \bar{\mathbf{H}}\left( \mathbf{x}^\circ\right)  \right)^{-1}=\left( \tilde{\mathbf{h}}_n\left( x_n^\circ\right)\tilde{\mathbf{h}}_n\left( x_n^\circ\right)^H+\tilde{\mathbf{H}}_n\left( x_n^\circ\right)\right) ^{-1}\nonumber\\
	&=\tilde{\mathbf{H}}_n\left( x_n^\circ\right)^{-1}-\frac{\tilde{\mathbf{H}}_n\left( x_n^\circ\right)^{-1}\tilde{\mathbf{h}}_n\left( x_n^\circ\right)\tilde{\mathbf{h}}_n\left( x_n^\circ\right)^H\tilde{\mathbf{H}}_n\left( x_n^\circ\right)^{-1}}{1+\tilde{\mathbf{h}}_n\left( x_n^\circ\right)^H\tilde{\mathbf{H}}_n\left( x_n^\circ\right)^{-1}\tilde{\mathbf{h}}_n\left( x_n^\circ\right)},
\end{align}
where ${{\tilde {\mathbf{H}}}_n}\left( x_n^\circ\right) = \sum\nolimits_{n' = 1,n' \ne n}^N {\tilde {\mathbf{h}}_{n'}\left( x_{n'}^\circ\right){{\tilde {\mathbf{h}}}_{n'}\left( x_{n'}^\circ\right)^H}} \in \mathbb{C}^{K \times K}$. By fixing the remaining PAs, the large-timescale position of PA $n$ can be obtained as
\begin{equation}\label{large_scale}
	x_n^\circ = \arg \mathop {\max }\limits_{x_n^\circ \in \tilde{\mathcal{S}}_n}  \frac{\tilde{\mathbf{h}}_n\left( x_n^\circ\right)^H\tilde{\mathbf{H}}_n\left( x_n^\circ\right)^{-2}\tilde{\mathbf{h}}_n\left( x_n^\circ\right)}{1+\tilde{\mathbf{h}}_n\left( x_n^\circ\right)^H\tilde{\mathbf{H}}_n\left( x_n^\circ\right)^{-1}\tilde{\mathbf{h}}_n\left( x_n^\circ\right)}.
\end{equation}
By alternately updating $\left\{x_n^\circ\right\}_{n=1}^N$ until convergence is reached, the large-timescale PA positions can be obtained.
\subsubsection{Small-Timescale PA Position Optimization}
After the large-timescale PA positions are determined, the PA positions are refined at the small timescale based on the instantaneous CSI to adapt to the NLoS channel variations in real time. Thus, the small-timescale PA position optimization problem is given by 
\begin{equation}\label{max5}
	\mathop {\mathrm{max} }\limits_{\mathbf{x}} \:R\left( {\mathbf{x}} \right) \quad\mathrm{s.t.} \: \eqref{c3}.
\end{equation}
To efficiently tackle problem \eqref{max5}, the projected gradient ascent algorithm is adopted. Specifically, the initial small-timescale PA position vector is set as $\mathbf{x}^{\left( 0\right) }=\mathbf{x}^\circ$, which is determined by the large-timescale PA positions. Let $\mathbf{x}^{\left( t-1\right) } = [ x_1^{\left( t-1\right)},\ldots,x_N^{\left( t-1\right)}] ^T$ represent the small-timescale PA position vector in the ($t-1$)-th iteration, where $t$ is the iteration index. Denote the gradient of $R\left( \mathbf{x}\right) $ with respect to $\mathbf{x}$ by ${\nabla_\mathbf{x}}R\left( \mathbf{x} \right)$, whose closed-form expression is derived in Appendix \ref{appendixB}. Then, the small-timescale PA position vector is updated in the $t$-th iteration as
\begin{equation}\label{position_update}
	\mathbf{x}^{\left( t\right) } = \mathcal{P}\left(  \mathbf{x}^{\left( t-1\right) } + \varpi^{\left( t\right)} \nabla_{\mathbf{x}}R\left( \mathbf{x}^{\left( t-1\right)}\right) \right),
\end{equation}
where $\mathcal{P}\left( \mathbf{x}\right) $ is a projection function that projects any element of $\mathbf{x}$ violating the feasible region defined by constraint \eqref{c3} onto the nearest boundary point \cite{boun}. The step size $\varpi^{\left( t\right)}$ is chosen by the backtracking line search to ensure a sufficient ascent of the objective value in problem \eqref{max5} while preventing unstable updates. For each iteration, the step size is initialized to a sufficiently large value, denoted by $\varpi^{\left( t\right)}=\varpi_0$. The candidate step size is then recursively shrunk by multiplying a factor $\delta \in \left( 0,1\right) $, i.e., $\varpi^{\left( t\right)}\leftarrow\delta \varpi^{\left( t\right)}$, until the Armijo-Goldstein condition is satisfied, which is given by
\begin{equation}\label{condition}
	R\left( \mathbf{x}^{\left(t \right) }\right)  \ge R\left( \mathbf{x}^{\left(t-1 \right)}\right)  + \xi \varpi^{\left(t \right)}\left\| \nabla_{\mathbf{x}} R\left( \mathbf{x}^{\left(t-1 \right)}\right) \right\|_2^2 ,
\end{equation}
where $\xi \in \left(0,1 \right) $ is a predefined constant that controls the required improvement of the objective value in problem \eqref{max5} at each iteration.
\subsection{Overall Algorithm}
\begin{algorithm}[!t]
	\caption{Proposed algorithm for solving problem \eqref{max1}}
	\label{alg2}
	\renewcommand{\algorithmicrequire}{\textbf{Input:}}
	\renewcommand{\algorithmicensure}{\textbf{Output:}}
	\begin{algorithmic}[1]
		\REQUIRE $N$, $K$, $S_\mathrm{x}$, $\lambda$, $h$, $n_\mathrm{e}$, $\sigma^2$, $P$, $\alpha$, $\beta_0$, $\{\mathbf{u}_k\}$, $\{y_n\}$, $\{\rho_n\}$, $\{L_{k,n}\}$, $\{\theta_{k,n}^l\}$, $\{b_{k,n}^l\}$, $\{D_n\}$, $\{\tilde{Q}_n\}$, $\{d_{k,n}\}$, $\{\kappa_{k,n}\}$, $\varpi_0$, $\delta$, $\xi$, $\epsilon$.
		\ENSURE $\mathbf{x}$, $\mathbf{W}$.\\
		\% \textit{Large-timescale PA position optimization}
		\REPEAT
		\FOR{$n = 1:1:N$}
		\STATE Update the large-timescale position $x_n^\circ$ of PA $n$ via \eqref{large_scale}.
		\ENDFOR
		\UNTIL{Decrease in the objective value of problem \eqref{max4} is less than $\epsilon$}\\
		\% \textit{Small-timescale PA position optimization}
		\STATE Initialize the small-timescale PA position vector as $\mathbf{x}^{\left(0 \right) }=\mathbf{x}^\circ$ and set $t = 0$.
		\STATE Calculate $R\left(\mathbf{x}^{\left(0 \right) } \right) $ via \eqref{R_small}.
		\REPEAT
		\STATE Set $t=t+1$.
		\STATE Calculate the gradient $\nabla_{\mathbf{x}} R\left( \mathbf{x}^{\left(t-1 \right)}\right)$ according to Appendix \ref{appendixB}.
		\STATE Initialize the step size as $\varpi^{\left( t\right)}=\varpi_0$.
		\STATE Update the small-timescale PA position vector $\mathbf{x}^{\left(t \right) }$ via \eqref{position_update}.
		\WHILE{condition \eqref{condition} is not satisfied}
		\STATE Shrink the step size $\varpi^{\left( t\right)}\leftarrow\delta \varpi^{\left( t\right)}$. 
		\STATE Update the small-timescale PA position vector $\mathbf{x}^{\left(t \right) }$ via \eqref{position_update}.
		\ENDWHILE
		\UNTIL{$R\left(\mathbf{x}^{\left(t \right) } \right) -R\left(\mathbf{x}^{\left(t-1 \right) } \right) < \epsilon$}
		\STATE Obtain the optimized small-timescale PA position vector $\mathbf{x}=\mathbf{x}^{\left(t\right)}$.
		\STATE Calculate the transmit beamforming matrix $\mathbf{W}$ via \eqref{beamforming}.
		\RETURN $\mathbf{x}$, $\mathbf{W}$.
	\end{algorithmic}
\end{algorithm}
The overall algorithm for solving problem \eqref{max1} is detailed in Algorithm \ref{alg2}. In lines 1-5, the large-timescale PA positions are alternately updated based on the statistical CSI until convergence is reached. Next, the small-timescale PA positions are initialized to the optimized large-timescale PA positions in line 6 and are iteratively optimized for each channel realization via the projected gradient ascent algorithm. For each iteration, the step size is determined by the backtracking line search in lines 13-16, and the algorithm terminates when the increase in the objective value of problem \eqref{max5} falls below a predefined threshold $\epsilon$. The complexity for optimizing the large-timescale PA positions in lines 1-5 is $\mathcal{O}(T_1K^2(NK+\sum\nolimits_{n=1}^N\tilde{Q}_n )   ) $, where $T_1$ denotes the maximum number of iterations for the large-timescale PA position optimization. The complexity for calculating the gradient in line 10 is $\mathcal{O}(K^3+NK^2+L_\mathrm{sum} ) $, where $L_\mathrm{sum}=\sum\nolimits_{k = 1}^K {\sum\nolimits_{n = 1}^N {{L_{k,n}}} } $. The complexity for updating the step size in lines 13-16 is $\mathcal{O}( T_2 (K^3 + N K^2 + L_\mathrm{sum}) )$, with $T_2$ being the maximum number of iterations for backtracking line search. Thus, the total computational complexity of Algorithm \ref{alg2} is $\mathcal{O}(T_1K^2(NK+\sum\nolimits_{n=1}^N\tilde{Q}_n ) + T_3T_2 (K^3 + N K^2 + L_\mathrm{sum})  ) $, where $T_3$ denotes the maximum number of iterations for the small-timescale PA position optimization.
\section{Simulation Results}\label{5}
In this section, simulation results are provided for the considered system. The simulation setup and benchmark schemes are first illustrated, and then the simulation results under single-user and multi-user scenarios are presented.
\subsection{Simulation Setup and Benchmark Schemes}
In the simulations, the spacing between adjacent waveguides along the $y$-axis is set to $\frac{S_\mathrm{y}}{N-1}$, and thus the $y$-axis coordinate of waveguide $n$ is given by $y_n=\frac{S_\mathrm{y}\left(n-1 \right) }{N-1}$. Without loss of generality, the Rician factor, the number of channel paths, the length of the small-timescale movement region, and the search resolution in Algorithm \ref{alg1} are assumed to be identical for all users and PAs, i.e., $\kappa_{k,n}\triangleq\kappa$, $L_{k,n}\triangleq L$, $D_n \triangleq D$, and $Q_n\triangleq Q$, $\forall k,n$. Unless otherwise specified, the default simulation parameters are listed in Table \ref{tab1}. The performance is evaluated by averaging the ergodic max-min rate over multiple realizations of user locations. Specifically, the users are distributed within a square area of size $S_\mathrm{x} \times S_\mathrm{y}$, where $S_\mathrm{x} = S_\mathrm{y}\triangleq S$, based on which the corresponding statistical CSI is obtained. Then, the ergodic max-min rate is obtained by averaging 100 instantaneous max-min rates, which is further averaged over 100 realizations of user locations. The results obtained by the proposed algorithms are referred to as \textit{Proposed}. In addition, the following benchmark schemes are considered for performance comparison:
\begin{itemize}
	\item \textit{Upper bound}: For the single-user case, the upper bound is approximated by increasing the search resolution $Q$ in Algorithm \ref{alg1} to $10^4$. For the multi-user case, the upper bound is given by \eqref{R_tup}.
	\item \textit{Large-timescale positioning only}: The PA positions are optimized only at the large timescale, without any small-timescale PA position refinement, which can be regarded as the conventional two-timescale optimization scheme for PASS \cite{PASS14}.
	\item \textit{Small-timescale movement only}: The large-timescale PA positions are fixed at the center of the waveguides, i.e., $x_n^\circ={S_\mathrm{x}}/{2}$, $\forall n$, and only small-timescale PA position refinement is performed.
	\item \textit{Sparse fixed-position antenna (FPA)}: The $N$ antennas form a sparse uniform linear array (ULA) parallel to the $y$-axis with an inter-antenna spacing of $\frac{S_\mathrm{y}}{N-1}$ and are deployed at the center of the square area. The position of antenna $n$ is given by $\left[ \frac{S_\mathrm{x}}{2},\frac{S_\mathrm{y}\left(n-1 \right) }{N-1},h\right]^T$.
	\item \textit{Dense FPA}: The $N$ antennas form a dense ULA parallel to the $y$-axis with an inter-antenna spacing of ${\lambda}/{2}$ and are deployed at the center of the square area. The position of antenna $n$ is given by $\left[ \frac{S_\mathrm{x}}{2},\frac{S_\mathrm{y}}{2}+\left( n-1-\frac{N-1}{2}\right)\frac{\lambda}{2} ,h\right]^T$.
\end{itemize}
\begin{table}[!t]
	\caption{Simulation Parameters}
	\label{tab1}
	\centering
	\renewcommand{\arraystretch}{1.2}
	\begin{tabular}{lll}
		\toprule
		\textbf{Parameter} & \textbf{Description} & \textbf{Value} \\
		\midrule
		$N$ & Number of waveguides & 16 \\
		$K$ & Number of users & 12 \\
		$S$ & Side length of square area & 100 m \\
		$L$ & Number of channel paths & 4 \\
		$\kappa$ & Rician factor of each user's channel & 10 dB \\
		$\beta_0$ & Average channel gain at reference distance & $-40$ dB \\
		$\alpha$ & Path loss exponent & 2.8 \\
		$h$ & Height of waveguides and PAs & 10 m \\
		$\lambda$ & Carrier wavelength & 0.01 m \\
		$n_\mathrm{e}$ & Effective refractive index & 1.4 \\
		$\rho_n$ & Power-scaling factor for each PA & 1 \\
		$P$ & Maximum transmit power & 10 dBm \\
		$\sigma^2$ & Average noise power & $-90$ dBm \\
		$D$ & Length of small-scale movement region & $5\lambda$ \\
		$Q$ & Search resolution in Algorithm \ref{alg1} & $10^{2}$ \\
		$\tilde{Q}_n$ & Search resolution in Algorithm \ref{alg2} & $10^{2}$ \\
		$\varpi_0$ & Initial step size in Algorithm \ref{alg2} & $10\lambda$ \\
		$\delta$ & Scaling factor in Algorithm \ref{alg2} & 0.5 \\
		$\xi$ & Control parameter in Algorithm \ref{alg2} & 0.1 \\
		$\epsilon$ & Predefined threshold in Algorithm \ref{alg2} & $10^{-5}$ \\
		\bottomrule
	\end{tabular}
\end{table}
\subsection{Single-User System}
\begin{figure}[!t]
	\centering
	\includegraphics[width=1\linewidth]{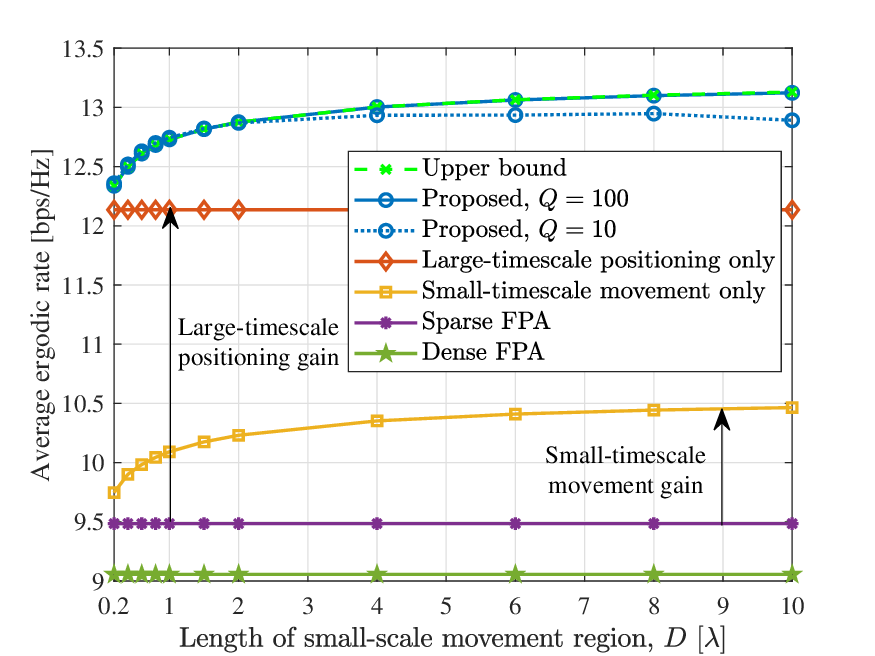}
	\caption{Average ergodic rate versus length of small-timescale movement region.}
	\label{sg_D}
\end{figure}
First, we consider the single-user setup, i.e., $K = 1$. Fig. \ref{sg_D} shows the average ergodic rate versus the length $D$ of the small-timescale movement region. As shown, the schemes that incorporate small-timescale PA movement, including the proposed two-timescale scheme and the small-timescale movement only scheme, achieve a monotonically increasing ergodic rate with increasing $D$. This is because the small-timescale PA position refinement can exploit the small-scale channel fading to enhance the effective channel gain. As $D$ becomes larger, each PA can explore a wider spatial region, which enables a more thorough exploitation of spatial DoFs. In contrast, the performance of schemes without small-timescale PA movement, such as the large-timescale positioning only scheme and the two FPA schemes, remains unchanged with respect to $D$. This behavior is expected since the large-timescale PA positioning is mainly used to shorten the signal propagation distance and thus reduce large-scale path loss. By jointly leveraging large-timescale and small-timescale antenna position adjustments, the proposed two-timescale scheme simultaneously captures the benefits of both positioning scales, thereby achieving a significantly higher ergodic rate. Moreover, the proposed scheme achieves a performance very close to the upper bound, which demonstrates that the parallel search strategy in Algorithm \ref{alg1} can effectively identify near-optimal positions for all PAs. Furthermore, when the search resolution $Q$ in Algorithm \ref{alg1} is reduced from 100 to 10, only a slight performance degradation is observed at large values of $D$, while the computational complexity of Algorithm \ref{alg1} is reduced by about one order of magnitude. This result highlights the favorable performance-complexity trade-off of the proposed scheme.

\begin{figure}[!t]
	\centering
	\includegraphics[width=1\linewidth]{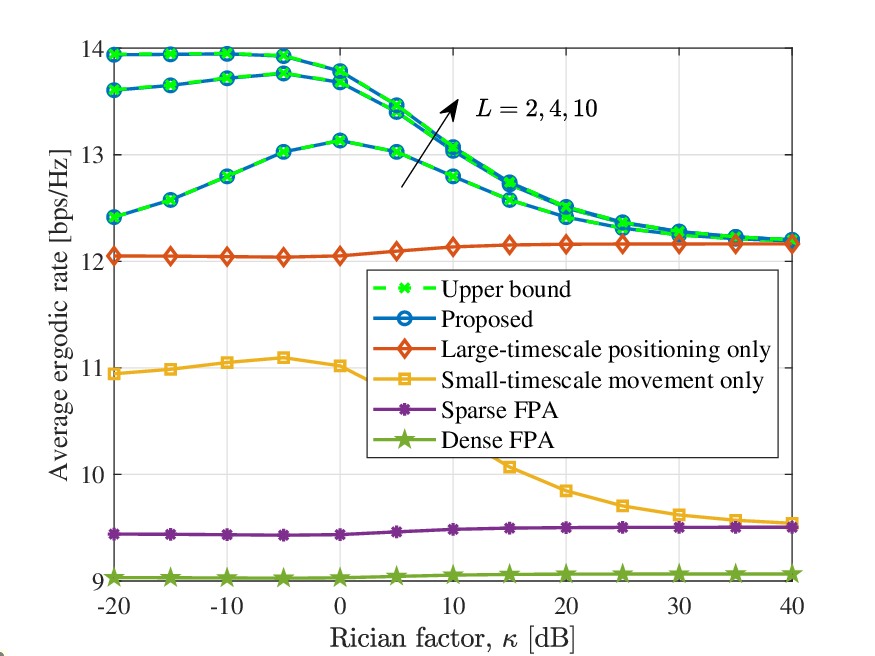}
	\caption{Average ergodic rate versus Rician factor.}
	\label{sg_kappa}
\end{figure}
Fig. \ref{sg_kappa} illustrates the average ergodic rate versus the Rician factor $\kappa$ for different numbers of channel paths $L$. When $L=2$, the proposed scheme exhibits a non-monotonic trend with respect to $\kappa$, where the ergodic rate first increases and then decreases, reaching its maximum around $\kappa=0$ dB. This is because, for very small or very large $\kappa$, the channel is effectively dominated by a single path when $L=2$, whereas at $\kappa=0$ dB, two paths with equal power coexist, which maximizes the multi-path diversity gain. As $L$ increases, this non-monotonic behavior gradually disappears, as a larger number of channel paths ensures sufficient diversity even when $\kappa$ is small. Moreover, when $\kappa$ becomes large, e.g., $\kappa=40$ dB, the performance of the proposed scheme approaches that of the large-timescale positioning only scheme, while the small-timescale movement only scheme achieves performance close to that of the sparse FPA scheme. This is because a large $\kappa$ corresponds to an almost pure LoS channel between each PA and the user, under which multi-path diversity is essentially absent. As a result, in the LoS scenario, the small-timescale PA movement becomes ineffective, and it is sufficient to rely on large-timescale PA positioning to place the PAs close to the user for path loss reduction, as discussed in Section \ref{pf}-1).
\subsection{Multi-User System}\label{mu_section}
Next, we consider the multi-user scenario, i.e., $K > 1$. To comprehensively evaluate the different schemes, the following user distributions are considered:
\begin{itemize}
	\item \textit{Random users}: The users are randomly distributed within the entire square area.
	\item \textit{Clustered users}: Three non-overlapping circular hotspot areas, each with a radius of 0.5 m, are randomly distributed within the square area. The numbers of users in each hotspot follow the ratio of 1:2:3, with users randomly distributed within their respective hotspot areas.
\end{itemize}

\begin{figure}[!t]
	\centering
	\subfloat[Objective value of PA position optimization]{\label{mu_conv}\includegraphics[width=0.5\columnwidth]{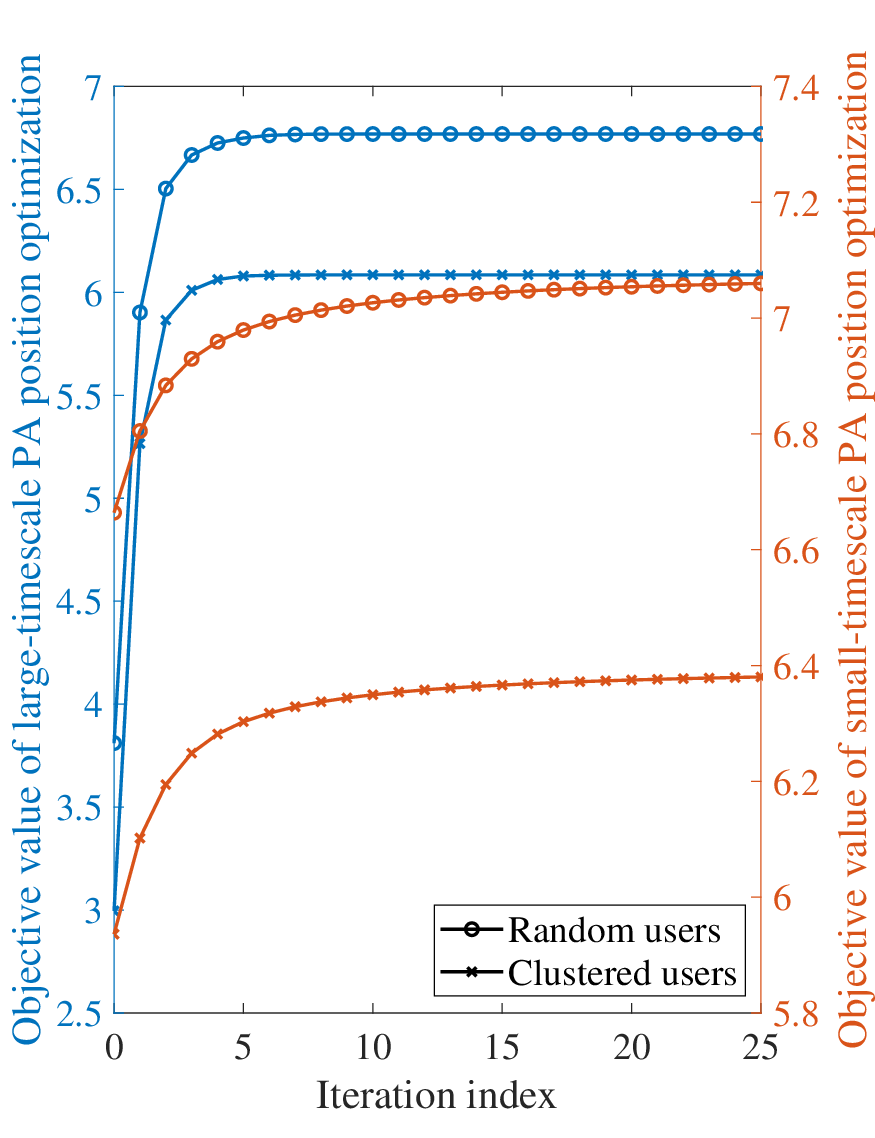}}
	\subfloat[Average channel gain and normalized channel cross-correlation]{\label{mu_conv_channel}\includegraphics[width=0.5\columnwidth]{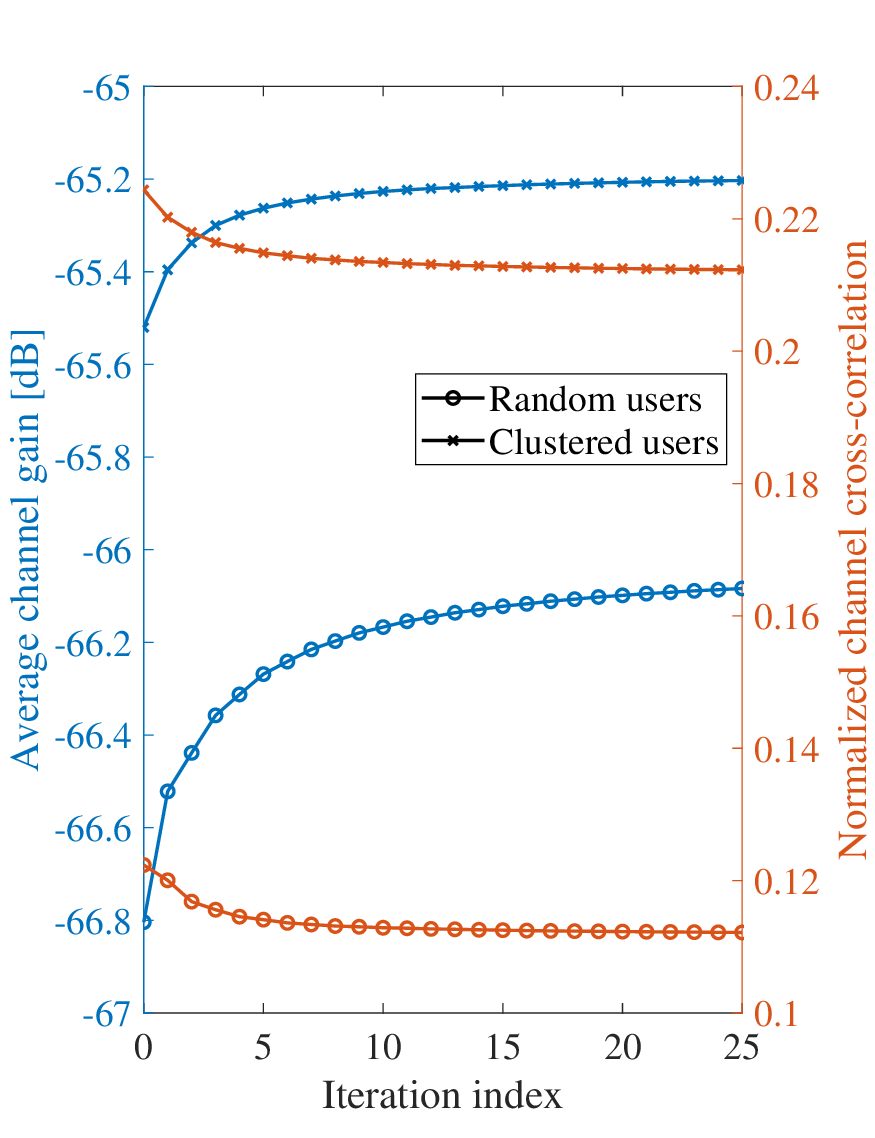}}
	\caption{Convergence behaviors of proposed algorithms.}
	\label{conv}
\end{figure}
Fig. \ref{conv} depicts the convergence behaviors of the proposed algorithms. Specifically, Fig. \ref{conv}\subref{mu_conv} plots the objective values of the large-timescale and small-timescale PA position optimizations versus the iteration index. For both random and clustered users, the objective values increase monotonically and stabilize within 25 iterations, which demonstrates the fast convergence of the proposed algorithms. In addition, the small-timescale PA position optimization converges more slowly than the large-timescale PA position optimization, since it relies on the more complicated instantaneous CSI. Moreover, the objective values for clustered users are consistently lower than those for random users. This is because users concentrated in hotspot areas exhibit stronger channel correlation, which leads to more severe inter-user interference and thus deteriorates the system performance. This conclusion can be further verified by Fig. \ref{conv}\subref{mu_conv_channel}, where the average channel gain and the normalized channel cross-correlation are shown. They are respectively defined as the expected values of $\frac{1}{K} \sum\nolimits_{1 \le k \le K} \left\| \mathbf{h}_k\left( \mathbf{x}\right)  \right\|_2^2$ and $\frac{1}{K\left( K-1\right) } \sum\nolimits_{1 \le k \ne k' \le K}
\frac{\left| \mathbf{h}_k\left( \mathbf{x}\right) ^H \mathbf{h}_{k'}\left( \mathbf{x}\right)  \right|}{\left\| \mathbf{h}_k\left( \mathbf{x}\right)  \right\|_2\left\| \mathbf{h}_{k'}\left( \mathbf{x}\right)  \right\|_2}$ over random channel realizations. It can be seen that the average channel gain increases with the iteration index, while the normalized channel cross-correlation decreases, confirming that the PA position optimization improves system performance by simultaneously enhancing the received signal power and mitigating inter-user interference. Although the clustered-user case achieves a higher channel gain in Fig. \ref{conv}\subref{mu_conv_channel}, it also yields a higher channel cross-correlation, and the resulting stronger inter-user interference explains its inferior objective values compared to the random-user case in Fig. \ref{conv}\subref{mu_conv}.

\begin{figure}[!t]
	\centering
	\includegraphics[width=1\linewidth]{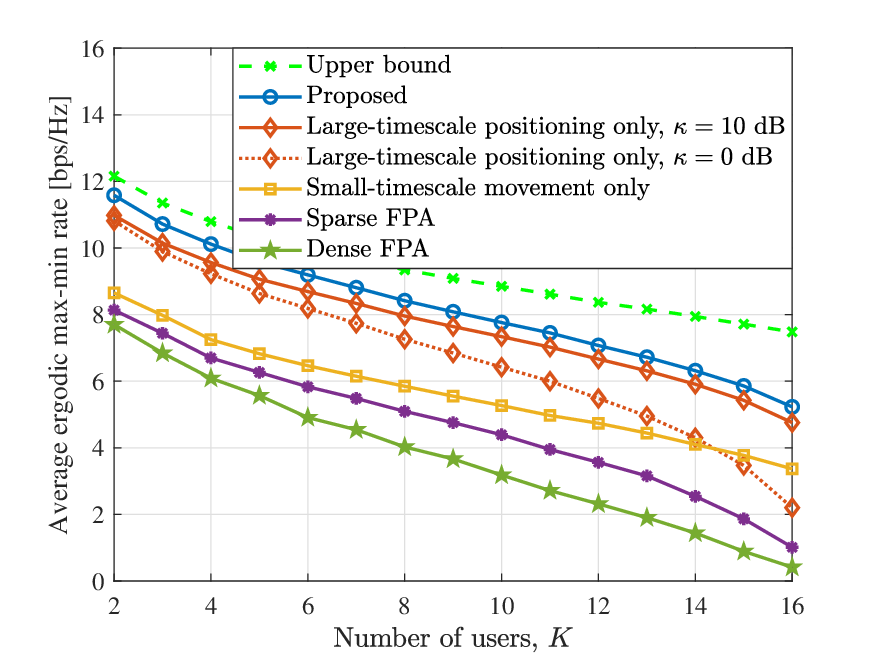}
	\caption{Average ergodic max-min rate versus number of users.}
	\label{mu_K}
\end{figure}
Fig. \ref{mu_K} compares the average ergodic max-min rates of different schemes versus the number of users $K$ under the random user distribution. As $K$ increases, the ergodic max-min rates of all schemes decrease monotonically, since a larger number of users leads to more severe inter-user interference, thereby degrading the minimum achievable rate under the total transmit power constraint. Consequently, the performance gap between the proposed scheme and the upper bound, obtained under the ideal assumption of no inter-user interference, gradually widens with increasing $K$. Compared to the other two PA schemes, the large-timescale positioning only scheme, which is designed based on statistical CSI, suffers more severe performance degradation in channels with rich multi-path components, e.g., $\kappa = 0$ dB, and is eventually outperformed by the small-timescale movement only scheme in the large-$K$ regime. This suggests that the small-timescale PA movement is essential for exploiting multi-path diversity in such scenarios. In contrast, in LoS-dominated channels, large-timescale PA positioning alone is sufficient to achieve satisfactory performance while avoiding the overhead of small-timescale PA movement. In addition, the performance gap between the small-timescale movement only scheme and the sparse FPA scheme increases with $K$, demonstrating that the small-timescale PA movement can effectively reduce channel correlation among users and becomes more beneficial in dense multi-user scenarios. 

\begin{figure}[!t]
	\centering
	\subfloat[Random users]{\label{mu_S_rand}\includegraphics[width=1\columnwidth]{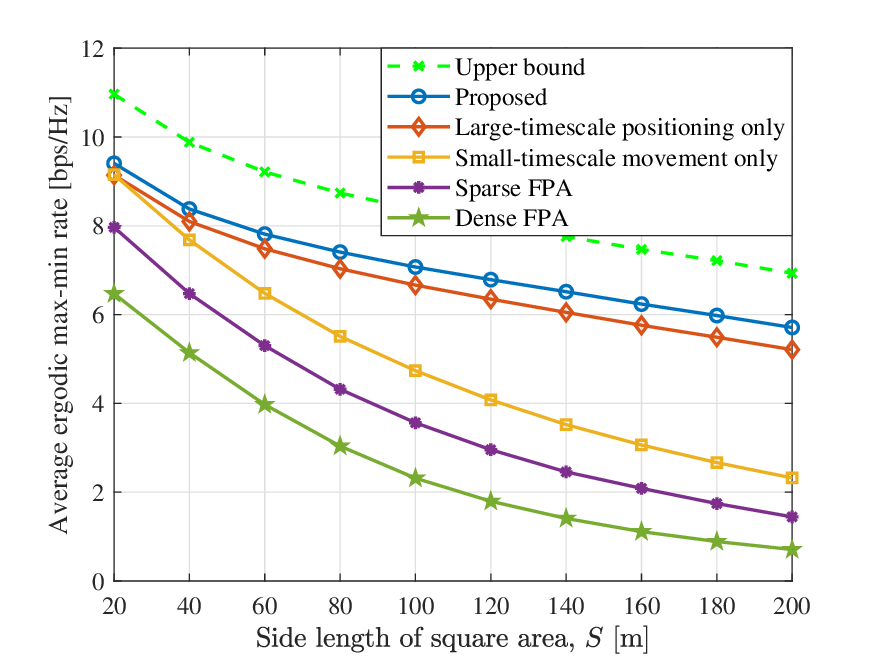}}\\
	\subfloat[Clustered users]{\label{mu_S_clu}\includegraphics[width=1\columnwidth]{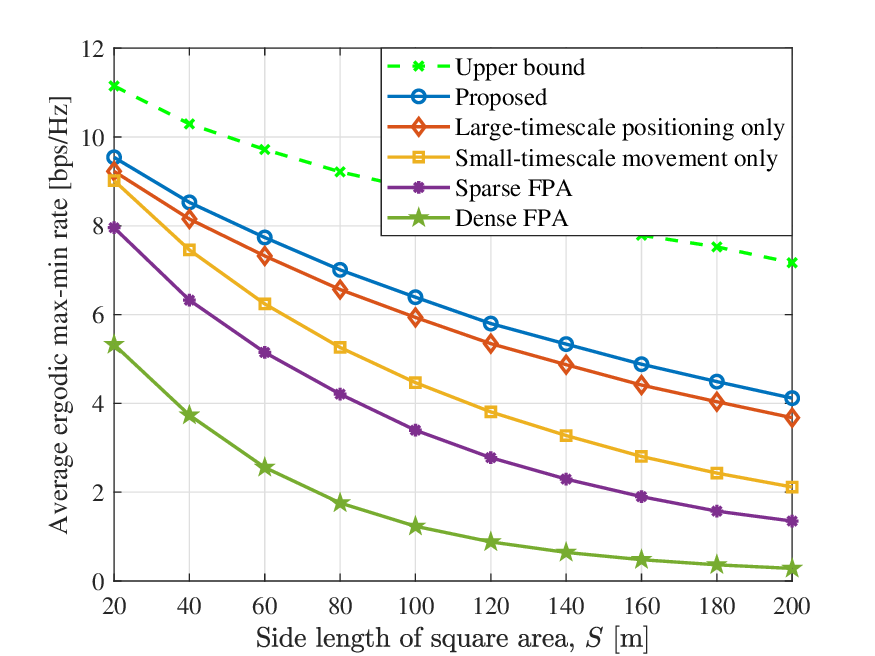}}
	\caption{Average ergodic max-min rate versus side length of square area.}
	\label{mu_S}
\end{figure}
Fig. \ref{mu_S} presents the average ergodic max-min rates for different values of the side length $S$ of the square area, under both random and clustered user distributions. As $S$ increases, the performance of all schemes degrades. The reason is as follows. For a fixed number of users, a larger $S$ implies a more dispersed user distribution. Under a fixed number of PAs, this dispersion increases the average distance between the PAs and the users, resulting in higher path loss and, consequently, a degradation in received signal power. It is worth noting that the proposed scheme and the large-timescale positioning only scheme exhibit much slower performance degradation compared to the small-timescale movement only scheme and the two FPA schemes. This is because the decentralized antenna array enabled by PASS fundamentally differs from the conventional centralized antenna deployment. By allowing PAs to move along the waveguides, antennas can be positioned closer to the users, thereby shortening the signal propagation distance and potentially alleviating blockage effects. As a result, large-timescale PA positioning plays a more critical role in improving system performance for PASS than small-timescale movement.

\begin{figure}[!t]
	\centering
	\includegraphics[width=1\linewidth]{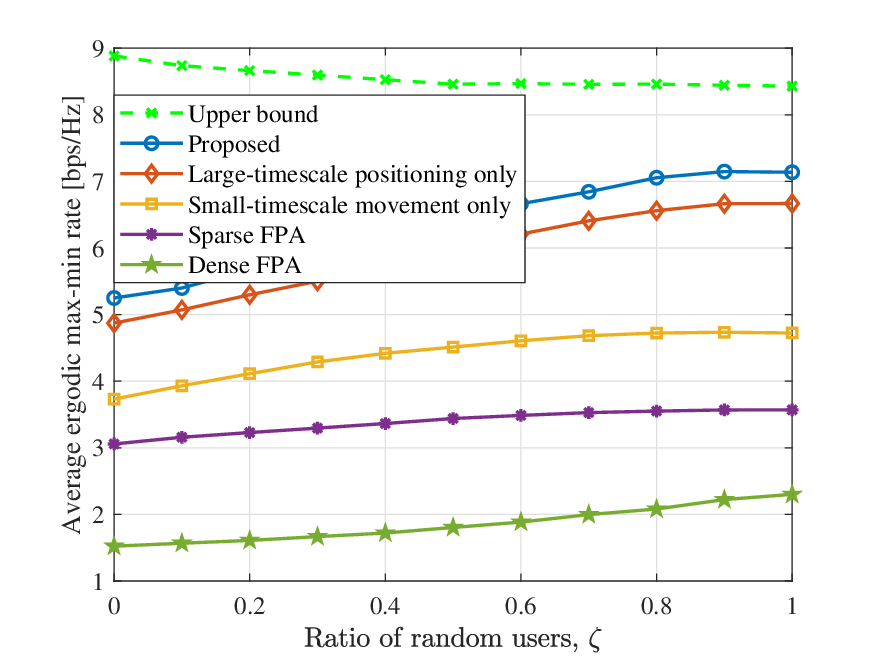}
	\caption{Average ergodic max-min rate versus ratio of random users.}
	\label{mu_ratio}
\end{figure}
To further illustrate the relationship between the ergodic max-min rate and the user distribution, we consider the case where a subset of users is clustered within a hotspot, while the remaining users are uniformly distributed over the entire square area. Specifically, a circular hotspot with a radius of 0.5 m is randomly generated inside the square area. Let $\zeta$ denote the ratio of random users, where each user is placed within the hotspot with probability $1-\zeta$ and is uniformly distributed outside the hotspot with probability $\zeta$. Fig. \ref{mu_ratio} shows the average ergodic max-min rates for different schemes as $\zeta$ varies. As $\zeta$ increases, the upper bound decreases, while the ergodic max-min rates of all other schemes increase. The reasons are explained as follows. On the one hand, increasing $\zeta$ reduces the degree of user clustering, which alleviates the inter-user interference. On the other hand, a more dispersed user distribution leads to larger average signal propagation distances, which increases large-scale path loss. These observations highlight the complementary roles of PA position adjustments at different timescales. For highly clustered users, small-timescale PA movement is particularly effective in reducing channel correlation. In contrast, when users are more randomly distributed, large-timescale PA positioning plays a dominant role by repositioning PAs closer to users to combat distance-dependent path loss. By jointly leveraging both mechanisms, the proposed two-timescale PA positioning scheme provides a robust solution that adapts well to diverse user distributions.

\begin{figure}[!t]
	\centering
	\includegraphics[width=1\linewidth]{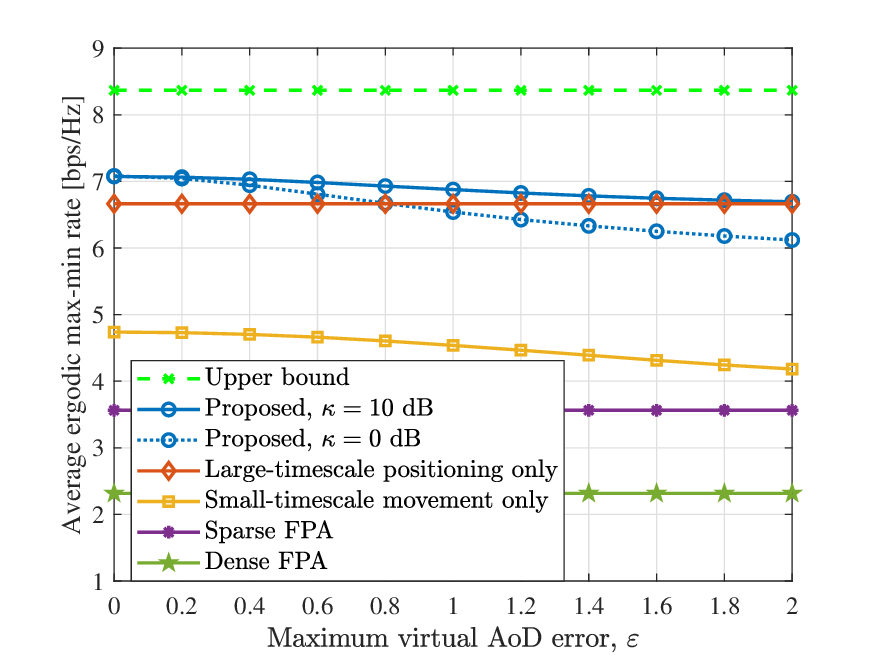}
	\caption{Average ergodic max-min rate versus maximum virtual AoD error.}
	\label{CSIerr}
\end{figure}
The above results are obtained under the assumption that the PASS has perfect knowledge of the CSI. However, in practical applications, acquiring instantaneous CSI requires the transmission/reception of training pilots, as well as CSI feedback between the transceivers. Due to channel fluctuations and noise, it is generally difficult to obtain perfect instantaneous CSI. Therefore, it is important to evaluate the impact of the CSI error on the performance of the proposed two-timescale PA positioning scheme. To this end, we define $\varepsilon$ as the maximum virtual AoD error to characterize the instantaneous CSI error. Specifically, the differences between the estimated and actual virtual AoDs are modeled as independent random variables following uniform distributions, i.e., $\vartheta_{k,n}^{l}-\hat{\vartheta}_{k,n}^{l}\sim \mathcal{U}\left[-\varepsilon/2,\varepsilon/2\right]$, $\forall k,n,l$, where $\vartheta_{k,n}^{l} $ and $\hat{\vartheta}_{k,n}^{l}$ denote the actual and estimated virtual AoDs, respectively. To focus on the impact of imperfect CSI on PA position optimization, the PA positions are optimized based on the estimated CSI, whereas the transmit beamforming matrix at the PASS is calculated using the actual channel vectors\footnote{In practice, after the PAs move to the optimized positions, the CSI at these fixed positions can be estimated using channel estimation methods for conventional FPA systems.}. Fig. \ref{CSIerr} shows the average ergodic max-min rates of different schemes versus the maximum virtual AoD error, where only the impact of CSI errors on PA position optimization is evaluated. It can be seen that the performance of the proposed scheme decreases slightly as $\varepsilon$ increases and gradually approaches that of the large-timescale positioning only scheme. This is because a larger instantaneous CSI error reduces the performance gain achieved by small-timescale PA movement, and this gain eventually diminishes when the estimated instantaneous CSI becomes sufficiently inaccurate. Moreover, the performance degradation becomes more pronounced as the Rician factor $\kappa$ decreases, indicating that accurate instantaneous CSI is particularly important in scenarios with rich NLoS propagation. Nevertheless, the proposed scheme consistently outperforms the benchmark schemes, demonstrating the robustness of the proposed PA position optimization against imperfect CSI.
\section{Conclusion}\label{6}
This paper investigated a two-timescale PA position optimization framework for PASS-enabled multi-user downlink communications. By decoupling PA deployment into large-timescale positioning over the entire waveguide using statistical CSI and small-timescale position refinement within a wavelength-scale region using instantaneous CSI, the proposed design avoids the impractical requirement of acquiring instantaneous CSI for arbitrary PA positions and achieves a favorable performance-complexity trade-off. For the single-user case, the optimal large-timescale PA positions were derived in closed form, and a parallel search algorithm was developed for efficient small-timescale position refinement to approach the derived upper bound. For the multi-user case, an upper bound on the max-min rate was analyzed under ideal interference elimination, and a low-complexity algorithm was proposed, where an element-wise search was used for large-timescale PA positioning and a projected gradient ascent method was used for small-timescale position refinement. Simulation results verified that the proposed scheme achieves performance close to the derived bounds and consistently outperforms benchmark schemes, highlighting that large-timescale position optimization mainly reduces large-scale path loss while small-timescale position refinement exploits small-scale channel fading to reduce inter-user interference, thereby providing a viable solution adaptable to diverse user distributions.
\appendices
\section{Proof of \eqref{pnset}}\label{appendixA}
The necessary and sufficient condition for achieving the upper bound \eqref{R_tup} is that the equalities in \eqref{orthogonal} and \eqref{gain} hold. When each PA has only a single path to each user, i.e., $L_{k,n}=1$, $\forall k,n$, the channel from the PASS to user $k$ can be simplified as
\begin{equation}\label{app_channel}
	\mathbf{h}_k(\mathbf{x})
	=
	\left[
	\begin{array}{c}
		\sqrt{{\beta_0}d_{k,1}^{-\alpha}}
		\rho_1 e^{\jmath \left(\frac{2\pi}{\lambda}\|\mathbf{p}_1-\mathbf{u}_k\|_2
			+\frac{2\pi}{\lambda_{\mathrm{W}}}x_1\right)}
		\\
		\vdots
		\\
		\sqrt{{\beta_0}d_{k,N}^{-\alpha}}
		\rho_N e^{\jmath \left(\frac{2\pi}{\lambda}\|\mathbf{p}_N-\mathbf{u}_k\|_2
			+\frac{2\pi}{\lambda_{\mathrm{W}}}x_N\right)}
	\end{array}
	\right]
	\in\mathbb{C}^{N\times 1}.
\end{equation}
Thus, the equality always holds in \eqref{gain}. The necessary and sufficient condition for satisfying the channel orthogonality \eqref{orthogonal} is derived as
\begin{align}
	&\mathbf{h}_k\left( {\mathbf{x}} \right)^H\mathbf{h}_{k'}\left( {\mathbf{x}} \right) = 0,\forall 1 \le k \ne k' \le K, \nonumber\\
	\mathop  \Leftrightarrow  \limits^{\left( {{\mathrm{d}_1}} \right)}& \sum\limits_{n = 1}^N {{\beta _0}\rho_n^2 \sqrt {\left( d_{k,n}d_{k',n}\right) ^{ - \alpha }} {e^{\jmath\frac{{2\pi }}{\lambda }\left( {{{\left\| {{\mathbf{p}_n} - {\mathbf{u}_{k'}}} \right\|}_2} - {{\left\| {{\mathbf{p}_n} - {\mathbf{u}_k}} \right\|}_2}} \right)}}}  = 0, \nonumber\\
	\mathop  \Leftrightarrow \limits^{\left( {{\mathrm{d}_2}} \right)} 
	&\left\{ 
	\begin{array}{l}
		{{{{\left\| {{\mathbf{p}_n} - {\mathbf{u}_{k'}}} \right\|}_2} - {{\left\| {{\mathbf{p}_n} - {\mathbf{u}_k}} \right\|}_2}}
			= \lambda \left( {z_{k,k',n} + \phi _{k,k',n}} \right)}, \\
		{\sum\limits_{n = 1}^N {{\beta _0}\rho_n^2\sqrt {\left( d_{k,n}d_{k',n}\right) ^{ - \alpha }} {e^{\jmath 2\pi {\phi _{k,k',n}}}}}  = 0 },
	\end{array} 
	\right.
\end{align}
where ($\mathrm{d}_1$) holds by substituting \eqref{app_channel} and ($\mathrm{d}_2$) holds because $e^{\jmath2\pi \left( z_{k,k',n} + \phi _{k,k',n} \right)} = {e^{\jmath 2\pi {\phi _{k,k',n}}}}$, $\forall z_{k,k',n} \in \mathbb{Z}$, $\forall 0 \le \phi_{k,k',n} < 1$. This thus completes the proof.
\section{Derivation of Gradient $\nabla_{\mathbf{x}} R\left( \mathbf{x}\right)$}\label{appendixB}
According to the chain rule and the definition in \eqref{R_small}, the gradient ${\nabla_\mathbf{x}}R\left( \mathbf{x} \right)$ is derived as
\begin{equation}
	{\nabla _\mathbf{x}}R\left( \mathbf{x} \right) = \left[ {\frac{{\partial R\left( \mathbf{x} \right)}}{{\partial {x_1}}}, \ldots ,\frac{{\partial R\left( \mathbf{x} \right)}}{{\partial {x_N}}}} \right]^T.
\end{equation}
Define $\mathbf{Z}\left(\mathbf{x} \right)=\mathbf{H}\left( \mathbf{x}\right)^H\mathbf{H}\left( \mathbf{x}\right)\in\mathbb{C}^{K\times K}$. The first-order partial derivative of $R\left( \mathbf{x} \right)$ with respect to $x_n$ is given by
\begin{align}
	&\frac{{\partial R\left( \mathbf{x} \right)}}{{\partial {x_n}}} = \frac{1}{{\left( {1 + \frac{P}{{\mathrm{tr}\left( {\mathbf{Z}{{\left( \mathbf{x} \right)}^{ - 1}}} \right){\sigma ^2}}}} \right)\ln 2}}\times\frac{\partial }{{\partial {x_n}}}\frac{P}{{\mathrm{tr}\left( {\mathbf{Z}{{\left( \mathbf{x} \right)}^{ - 1}}} \right){\sigma ^2}}} \nonumber\\
	&=   \frac{-P}{{\mathrm{tr}\left( {\mathbf{Z}{{\left( \mathbf{x} \right)}^{ - 1}}} \right)\left( {\mathrm{tr}\left( {\mathbf{Z}{{\left( \mathbf{x} \right)}^{ - 1}}} \right){\sigma ^2} + P} \right)\ln 2}}\mathrm{tr}\left( {\frac{{\partial \mathbf{Z}{{\left( \mathbf{x} \right)}^{ - 1}}}}{{\partial {x_n}}}} \right)\nonumber\\
	&= \frac{P}{{\mathrm{tr}\left( {\mathbf{Z}{{\left( \mathbf{x} \right)}^{ - 1}}} \right)\left( {\mathrm{tr}\left( {\mathbf{Z}{{\left( \mathbf{x} \right)}^{ - 1}}} \right){\sigma ^2} + P} \right)\ln 2}}  \nonumber\\
	& \hspace{1em}\times \mathrm{tr}\left( {{{\left( {\mathbf{Z}{{\left( \mathbf{x} \right)}^*}} \right)}^{ - 1}}\frac{{\partial \mathbf{Z}\left( \mathbf{x} \right)}}{{\partial {x_n}}}{{\left( {\mathbf{Z}{{\left( \mathbf{x} \right)}^*}} \right)}^{ - 1}}} \right),
\end{align}
where the element of ${{\partial \mathbf{Z}\left( \mathbf{x} \right)}}/{{\partial {x_n}}}$ in the $i$-th row and $j$-th column ($1 \le i,j \le K$) is derived as
\begin{align}
	&{\left[ {\frac{{\partial \mathbf{Z}\left( \mathbf{x} \right)}}{{\partial {x_n}}}} \right]_{i,j}} = \frac{{\partial \left( {\mathbf{f}_i}{{\left( \mathbf{x} \right)}^H}{\mathbf{B}_i}\mathbf{B}_j^H{\mathbf{f}_j}\left( \mathbf{x} \right)\right) }}{{\partial {x_n}}} \nonumber\\
	&= {\left( {\frac{{\partial {\mathbf{f}_i}\left( \mathbf{x} \right)}}{{\partial {x_n}}}} \right)^H}{\mathbf{B}_i}\mathbf{B}_j^H{\mathbf{f}_j}\left( \mathbf{x} \right) + {\mathbf{f}_i}{\left( \mathbf{x} \right)^H}{\mathbf{B}_i}\mathbf{B}_j^H\frac{{\partial {\mathbf{f}_j}\left( \mathbf{x} \right)}}{{\partial {x_n}}}.
\end{align}
Based on the definition in \eqref{fk}, the first-order partial derivative of ${\mathbf{f}_k}\left( \mathbf{x} \right)$  with respect to $x_n$ is given by
\begin{equation}
	\frac{{\partial {\mathbf{f}_k}\left( \mathbf{x} \right)}}{{\partial {x_n}}} = {\left[ {\mathbf{0}_{{L_1}}^T,{{\left( {\frac{{\partial \left( {g_n\left( {{x_n}} \right){\mathbf{f}_{k,n}}\left( {{x_n}} \right)} \right)}}{{\partial {x_n}}}} \right)}^T},\mathbf{0}_{{L_2}}^T} \right]^T},
\end{equation}
where ${L_1}=\sum\nolimits_{i = 1}^{n - 1} {{L_{k,i}}} $, ${L_2}=\sum\nolimits_{i = n+1}^{N} {{L_{k,i}}} $, and the $l$-th ($1\le l \le L_{k,n}$) element of ${{{\partial \left( {g_n\left( {{x_n}} \right){\mathbf{f}_{k,n}}\left( {{x_n}} \right)} \right)}}/{{\partial {x_n}}}}$ is derived as
\begin{align}
	\left[ {\frac{{\partial \left( {g_n\left( {{x_n}} \right){\mathbf{f}_{k,n}}\left( {{x_n}} \right)} \right)}}{{\partial {x_n}}}}\right]_l&=\jmath2\pi \left( {\frac{{\cos \theta _{k,n}^l}}{\lambda } + \frac{1}{{{\lambda _\mathrm{W}}}}} \right) \nonumber\\
	&\hspace{1em}\times \rho_n{e^{\jmath \left( {\frac{{2\pi }}{\lambda }\cos \theta _{k,n}^l\left( {{x_n} - x_n^\circ} \right) + \frac{{2\pi }}{{{\lambda _\mathrm{W}}}}{x_n}} \right)}}. 
\end{align}

\begin{IEEEbiography}[{\includegraphics[width=1in,height=1.25in,clip,keepaspectratio]{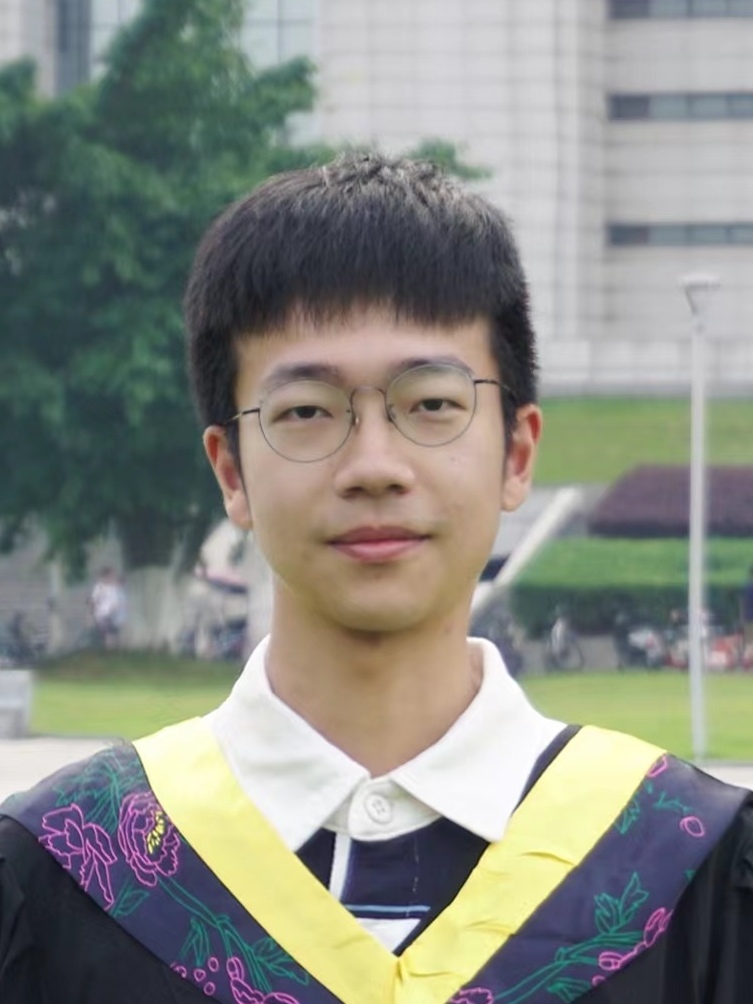}}]
	{Jingze Ding}(Graduate Student Member, IEEE) received the B.E. degree in information engineering from the University of Electronic Science and Technology of China (UESTC), Chengdu, China, in 2022. He is currently pursuing the Ph.D. degree with the School of Electronics, Peking University, Beijing, China. He is also a visiting Ph.D. student with the Department of Electrical and Computer Engineering, National University of Singapore, Singapore. His research interests include flexible antenna and optimization techniques.
\end{IEEEbiography}

\begin{IEEEbiography}[{\includegraphics[width=1in,height=1.25in,clip,keepaspectratio]{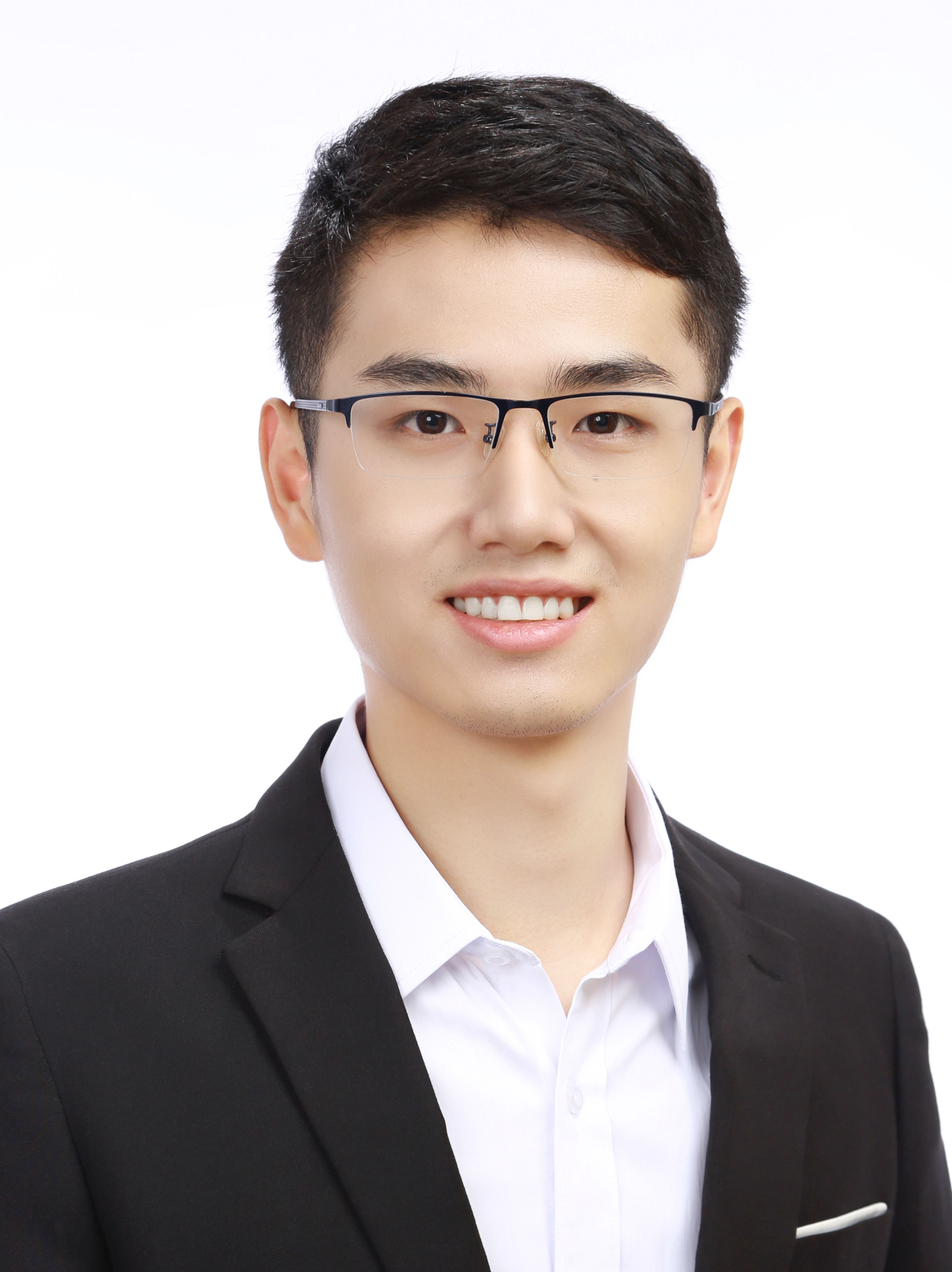}}]
	{Zijian Zhou}(Member, IEEE) received the B.Eng. degree in information engineering from Soochow University, Suzhou, China, in 2020. He is currently pursuing the Ph.D. degree with the School of Science and Engineering, The Chinese University of Hong Kong, Shenzhen, China. From 2020 to 2024, he was a Research Assistant with the School of Electronics, Peking University, Beijing, China. His research interests include polarforming for wireless communications, movable antenna (MA)-enabled communication systems, and convex optimization. He was honored as a Chun-Tsung scholar of Soochow University in 2019.
\end{IEEEbiography}

\begin{IEEEbiography}[{\includegraphics[width=1in,height=1.25in,clip,keepaspectratio]{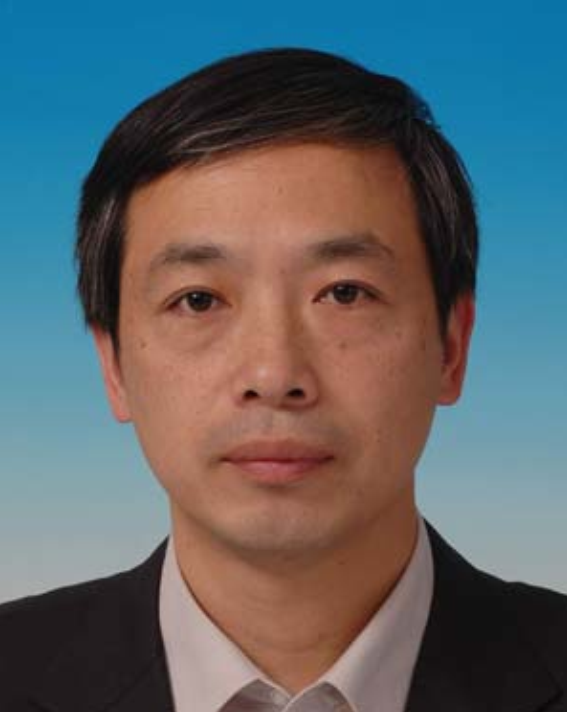}}]
	{Bingli Jiao}(Senior Member, IEEE) received the B.S. and M.S. degrees from Peking University, Beijing, China, in 1983 and 1988, respectively, and the Ph.D. degree from Saarland University, Saarbr\"ucken, Germany, in 1995. He became an Associate Professor in 1995 and a Professor with Peking University in 2000. He is currently the Director of the Joint Laboratory for Advanced Communication Research between Peking University and Geespace. He is a pioneer of co-frequency and co-time full-duplex as found in his early patent in 2006. His research interests include full-duplex communications, information theory, and signal processing.
\end{IEEEbiography}

\begin{IEEEbiography}[{\includegraphics[width=1in,height=1.25in,clip,keepaspectratio]{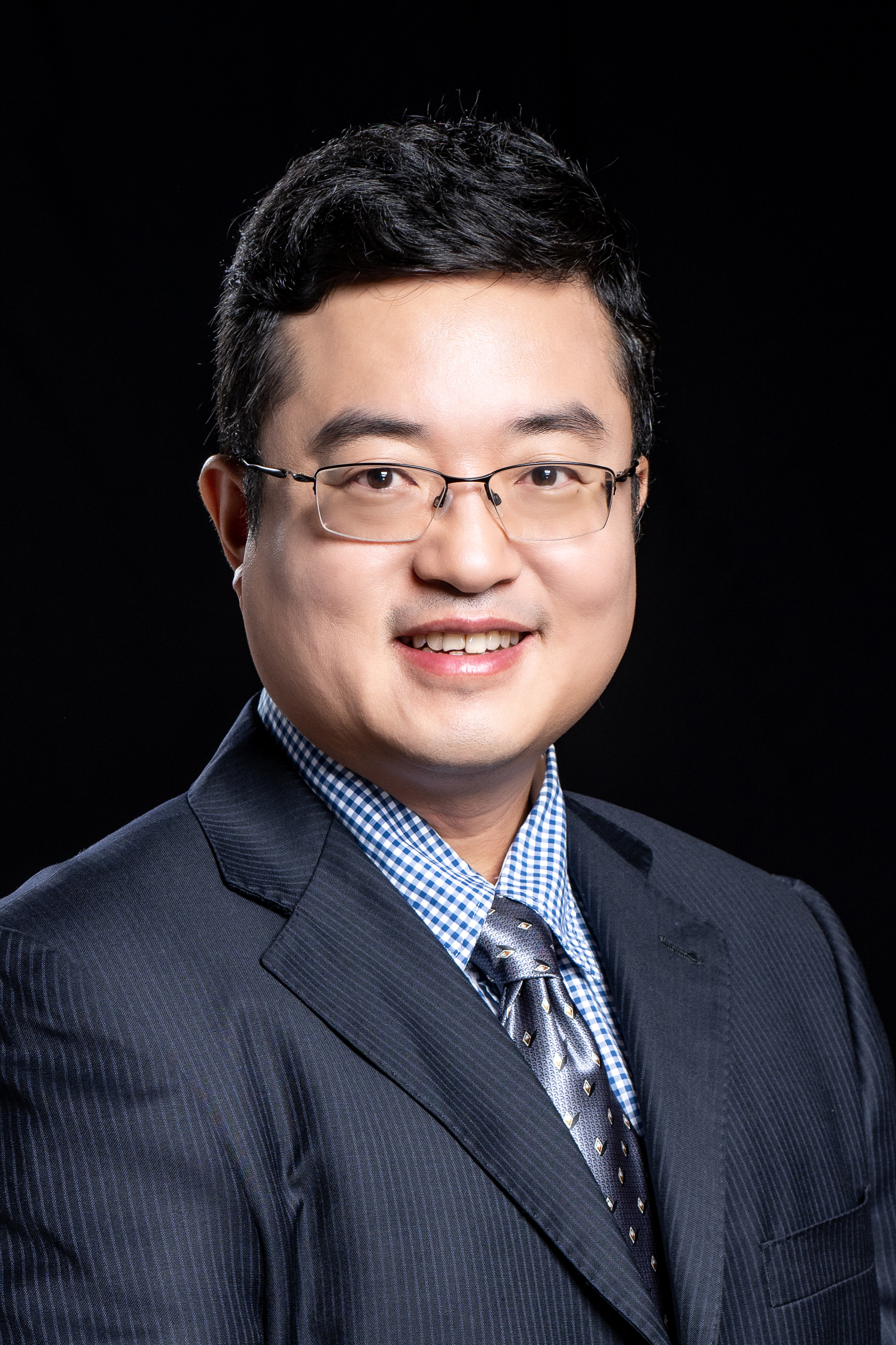}}]
	{Rui Zhang}(Fellow, IEEE) received the B.Eng. (first-class Hons.) and M.Eng. degrees from the National University of Singapore, Singapore, and the Ph.D. degree from the Stanford University, Stanford, CA, USA, all in electrical engineering.
	
	From 2007 to 2009, he worked as a research scientist at the Institute for Infocomm Research, ASTAR, Singapore. In 2010, he joined the Department of Electrical and Computer Engineering of National University of Singapore, where he is now a Provost's Chair Professor. He is also an Adjunct Professor with the School of Science and Engineering, The Chinese University of Hong Kong, Shenzhen, China. He has published over 600 papers, all in the field of wireless communications and networks. He has been listed as a Highly Cited Researcher by Thomson Reuters/Clarivate Analytics since 2015. His current research interests include intelligent surfaces, reconfigurable antennas, radio mapping, non-terrestrial communications, wireless power transfer, AI and optimization methods.      
	
	He was the recipient of the 6th IEEE Communications Society Asia-Pacific Region Best Young Researcher Award in 2011, the Young Researcher Award of National University of Singapore in 2015, the Wireless Communications Technical Committee Recognition Award in 2020, the IEEE Signal Processing and Computing for Communications (SPCC) Technical Recognition Award in 2021, and the IEEE Communications Society Technical Committee on Cognitive Networks (TCCN) Recognition Award in 2023. His works received 18 IEEE Best Journal Paper Awards, including the IEEE Marconi Prize Paper Award in Wireless Communications in 2015 and 2020, the IEEE Signal Processing Society Best Paper Award in 2016, the IEEE Communications Society Heinrich Hertz Prize Paper Award in 2017, 2020 and 2022, the IEEE Communications Society Stephen O. Rice Prize in 2021, etc. He served for over 30 international conferences as the TPC co-chair or an organizing committee member. He was an elected member of the IEEE Signal Processing Society SPCOM Technical Committee from 2012 to 2017 and SAM Technical Committee from 2013 to 2015. He served as the Vice Chair of the IEEE Communications Society Asia-Pacific Board Technical Affairs Committee from 2014 to 2015, a member of the Steering Committee of the IEEE Wireless Communications Letters from 2018 to 2021, a member of the IEEE Communications Society Wireless Communications Technical Committee (WTC) Award Committee from 2023 to 2025. He was a Distinguished Lecturer of IEEE Signal Processing Society and IEEE Communications Society from 2019 to 2020. He served as an Editor for several IEEE journals, including the IEEE TRANSACTIONS ON WIRELESS COMMUNICATIONS from 2012 to 2016, the IEEE JOURNAL ON SELECTED AREAS IN COMMUNICATIONS: Green Communications and Networking Series from 2015 to 2016, the IEEE TRANSACTIONS ON SIGNAL PROCESSING from 2013 to 2017, the IEEE TRANSACTIONS ON GREEN COMMUNICATIONS AND NETWORKING from 2016 to 2020, and the IEEE TRANSACTIONS ON COMMUNICATIONS from 2017 to 2022. He now serves as an Editorial Board Member of npj Wireless Technology, and the Chair of the IEEE Communications Society Wireless Communications Technical Committee (WTC) Award Committee. He is a Fellow of the Academy of Engineering Singapore.
\end{IEEEbiography}
\end{document}